\documentclass[12pt,a4paper]{article}

\usepackage{amsmath}
\usepackage{amsfonts}
\usepackage{amssymb}
\usepackage{graphicx}
\usepackage{color}
\usepackage{booktabs}
\usepackage{inputenc}
\usepackage[T1]{fontenc}
\usepackage{mathrsfs}
\usepackage{enumerate}
\usepackage{xcolor,url}
\usepackage{cancel,dsfont} 
\usepackage{braket}

\newcommand{\eea}{\end{eqnarray}}
\newcommand{\bea}{\begin{eqnarray}}

\def\be{\begin{equation}}
\def\ee{\end{equation}}

\def\nn{\nonumber}

\newcommand{\al}{\alpha}
\newcommand{\bt}{\beta}

\def\hinvMpc{h\,{\rm Mpc}^{-1}}
\def\Mpcinvh{{\rm Mpc}/h}

\newcommand{\kmax}{k_{\rm max }}

\newcommand{\code}[1]{\texttt{#1}}

\def\beq{\begin{equation}}
\def\eeq{\end{equation}}
\def\be{\begin{equation}}
\def\ee{\end{equation}}
\def\bea{\begin{eqnarray}}
\def\eea{\end{eqnarray}}

\def\nn{\nonumber}

\def\k{{\bf k}}

\newcommand{\ta}{\tilde{a}}

\newcommand{\mG}{\mathcal{G}}

\newcommand{\mU}{\mathcal{U}}
\newcommand{\mV}{\mathcal{V}}

\newcommand{\q}{\vec{q}}
\newcommand{\p}{\vec{p}}

\usepackage[colorlinks,bookmarks]{hyperref}
\definecolor{linkblue}{rgb}{0,0,0.8}
\definecolor{linkgreen}{rgb}{0,0.5,0}

\hypersetup{pdfpagemode=UseNone, pdfstartview=FitH, linkcolor=linkblue,
            citecolor=linkgreen, urlcolor=linkblue}

 \font\BF=cmmib10
\def\k{{\hbox{\BF k}}}
\def\q{{\hbox{\BF q}}}
\def\z{{\hbox{\BF z}}}
\def\s{{\hbox{\BF s}}}
\def\p{{\hbox{\BF p}}}
\def\r{{\hbox{\BF r}}}
\def\x{{\hbox{\BF x}}}

\begin{document}

\begin{center}

{\Large \bf {BOSS full-shape analysis\\[0.3cm]
 from the EFTofLSS with exact time dependence
}
}
\\[0.7cm]

{\large  Pierre Zhang${}^{1,2,3}$, Yifu Cai${}^{1,2,3}$  \\[0.7cm]}

\end{center}

\begin{center}

\vspace{.0cm}

{\normalsize { \sl $^{1}$ Department of Astronomy, School of Physical Sciences, \\
University of Science and Technology of China, Hefei, Anhui 230026, China}}\\
\vspace{.3cm}

{\normalsize { \sl $^{2}$ CAS Key Laboratory for Research in Galaxies and Cosmology, \\
University of Science and Technology of China, Hefei, Anhui 230026, China}}\\
\vspace{.3cm}

{\normalsize { \sl $^{3}$ School of Astronomy and Space Science, \\
University of Science and Technology of China, Hefei, Anhui 230026, China}}\\
\vspace{.3cm}

\vspace{.3cm}

\end{center}

\hrule \vspace{0.3cm}
{\small  \noindent \textbf{Abstract} 
We re-analyze the full shape of BOSS galaxy two-point function from the Effective-Field Theory of Large-Scale Structure at the one loop within $\Lambda$CDM with massive neutrinos using a big bang nucleosynthesis (BBN) prior, removing the Einstein-de Sitter (EdS) approximation in the time dependence of the loop, and, properly accounting for the redshift selection over the BOSS samples instead of assuming an effective redshift. 
We constrain, at $68\%$-confidence level (CL), the present-day matter fraction to $\Omega_m=0.322 \pm 0.018$, the Hubble constant to $H_0=69.1\pm 0.14$ (km/s)/Mpc, the $\log$-amplitude of the primordial spectrum to $\ln (10^{10} A_s) = 2.97 \pm 0.25$, the spectral tilt to $n_s = 0.938 \pm 0.082$, and bound the total neutrino mass to $<1.1$ at $95\%$-CL. 
We find no significant shift in the posteriors of the cosmological parameters due to the EdS approximation, but a marginal difference in $\ln (10^{10} A_s)$ due to the effective redshift approximation of about $0.4\sigma$, where $\sigma$ is the $68\%$-confidence interval. 
Regarding the EdS approximation, we check that the same conclusion holds on simulations of volume like DESI in $\Lambda$CDM and $w$CDM, with a BBN prior. 
In contrast, for an approximate, effective redshift, to be assumed, we advocate systematic assessments on redshift selection for ongoing and future large-volume surveys.

%\\[0.0cm]
\noindent

\vspace{0.3cm}}
\hrule

\vspace{0.3cm}
\newpage

\tableofcontents

\section{Introduction \label{sec:intro}}

\paragraph{Context} Recently, all parameters in $\Lambda$CDM have been measured with only a big bang nucleosynthesis~(BBN) prior on the baryon abundance by analyzing the full shape (FS) of the BOSS galaxy power spectrum (PS) or correlation function (CF) using the Effective Field Theory of Large-scale Structure (EFTofLSS) at one-loop order in~\cite{DAmico:2019fhj,Ivanov:2019pdj,Colas:2019ret,Zhang:2021yna}, and together with the tree-level bispectrum in~\cite{DAmico:2019fhj} (see also~\cite{Philcox:2020xbv} for other prior choices).
Extensions such as neutrino masses, effective number of relativistic species, smooth or clustering quintessence, or curvature, have also been bounded or measured from the BOSS FS using the EFTofLSS in combination with the reconstructed PS from BOSS and baryon acoustic oscillations (BAO) from eBOSS, as well as with supernovae redshift-distance relationship or cosmic microwave background~(CMB)  measurements~\cite{DAmico:2019fhj,Colas:2019ret,Zhang:2021yna,Ivanov:2019hqk,Philcox:2020vvt,DAmico:2020kxu,DAmico:2020tty,Chudaykin:2020ghx} (see also~\cite{Chen:2021wdi} for another recent analysis of BOSS FS and reconstructed CF). 
These limits are competitive with other probes: in particular, with BOSS FS and BBN only, the constraints on the Hubble constant $H_0$ and the present-day matter fraction $\Omega_m$ are comparable to the ones obtained by Planck~\cite{Planck:2018vyg}. 
Such precision could be achieved with current spectroscopic surveys since the degeneracies in the FS analysis are different than in the CMB. 
Additionally, it has been shown that more cosmological information can be extracted from the FS once redshift-space distortions are mitigated~\cite{DAmico:2021ymi,Ivanov:2021haa}. 
Besides, as the measurements from BOSS are independent of Planck, the FS analysis can help constrain models invented to alleviate the Hubble tension~\cite{DAmico:2020ods,Ivanov:2020ril,Niedermann:2020qbw,Smith:2020rxx}. 

\paragraph{EFTofLSS} All these results were obtained using the EFTofLSS, that has revealed to be a powerful tool to extract cosmological information from redshift surveys. 
We here provide a selected overview of its important advancements, that allowed to bring the framework to the level where it could be applied to the data. 
The initial formulation of the EFTofLSS was performed in Eulerian space in~\cite{Baumann:2010tm,Carrasco:2012cv}, and subsequently extended to Lagrangian space in~\cite{Porto:2013qua}. 
The dark matter power spectrum has been computed at one-, two- and three-loop orders in~\cite{Carrasco:2012cv, Carrasco:2013sva, Carrasco:2013mua, Carroll:2013oxa, Senatore:2014via, Baldauf:2015zga, Foreman:2015lca, Baldauf:2015aha, Cataneo:2016suz, Lewandowski:2017kes,Konstandin:2019bay}.
These calculations were accompanied by some  theoretical developments of the EFTofLSS, such as a careful understanding of renormalization~\cite{Carrasco:2012cv,Pajer:2013jj,Abolhasani:2015mra} (including rather-subtle aspects such as lattice-running~\cite{Carrasco:2012cv} and a better understanding of the velocity field~\cite{Carrasco:2013sva,Mercolli:2013bsa}), of several ways for extracting the value of the counterterms from simulations~\cite{Carrasco:2012cv,McQuinn:2015tva}, and of the non-locality in time of the EFTofLSS~\cite{Carrasco:2013sva, Carroll:2013oxa,Senatore:2014eva}.
These theoretical explorations also include an enlightening study in 1+1 dimensions~\cite{McQuinn:2015tva,Pajer:2017ulp}.
An IR-resummation of the long displacement fields had to be performed in order to reproduce the Baryon Acoustic Oscillation (BAO) peak, giving rise to the so-called IR-Resummed EFTofLSS~\cite{Senatore:2014vja,Baldauf:2015xfa,Senatore:2017pbn,Lewandowski:2018ywf,Blas:2016sfa}.
An account of baryonic effects was presented in~\cite{Lewandowski:2014rca,Braganca:2020nhv}. The dark-matter bispectrum has been computed at one- and two-loop in~\cite{Angulo:2014tfa, Baldauf:2014qfa, Baldauf:2021zlt}, the one-loop trispectrum in~\cite{Bertolini:2016bmt}, and the displacement field in~\cite{Baldauf:2015tla}.
The lensing power spectrum has been computed at two loops in~\cite{Foreman:2015uva}.
Biased tracers, such as halos and galaxies, have been studied in the context of the EFTofLSS in~\cite{ Senatore:2014eva, Mirbabayi:2014zca, Angulo:2015eqa, Fujita:2016dne, Perko:2016puo, Nadler:2017qto} (see also~\cite{McDonald:2009dh}), the halo and matter power spectra and bispectra (including all cross correlations) in~\cite{Senatore:2014eva, Angulo:2015eqa}. Redshift space distortions have been developed in~\cite{Senatore:2014vja, Lewandowski:2015ziq,Perko:2016puo}.
Neutrinos have been included in the EFTofLSS in~\cite{Senatore:2017hyk,deBelsunce:2018xtd}, clustering dark energy in~\cite{Lewandowski:2016yce,Lewandowski:2017kes,Cusin:2017wjg,Bose:2018orj}, and primordial non-Gaussianities in~\cite{Angulo:2015eqa, Assassi:2015jqa, Assassi:2015fma, Bertolini:2015fya, Lewandowski:2015ziq, Bertolini:2016hxg}.
Faster evaluation schemes for the calculation of some of the loop integrals have been developed in~\cite{Simonovic:2017mhp}.
Comparison with high-fidelity $N$-body simulations showing that the EFTofLSS can accurately recover the cosmological parameters have been performed in~\cite{DAmico:2019fhj,Colas:2019ret,Zhang:2021yna,Nishimichi:2020tvu,Chen:2020zjt}.

\paragraph{Motivation and Plan} With the advent of ongoing and future surveys with increasingly bigger volumes such as DESI~\cite{Aghamousa:2016zmz} or Euclid~\cite{Amendola:2012ys}, and the additional cosmological information that the EFTofLSS enables us to extract from the data, it is worthwhile to investigate all possible sources of systematics in the modeling and in the measurements. 
Until now, the FS analysis has been performed assuming the following two `time' approximations. 
\emph{One}, the time dependence in the loop is approximated by powers of the growth factor~\footnote{Note however the exception of~\cite{DAmico:2020tty}, where the BOSS FS has been analyzed using the exact time dependence in quintessence models. }. 
This is frequently called the Einstein-de Sitter (EdS) approximation. 
\emph{Two}, the evaluation of the FS is performed at a single effective redshift to account for the redshift selection over the observational samples. 
The goal of this paper is to investigate the impact of these two approximations on the determination of the cosmological parameters. 
Recently in~\cite{Donath:2020abv} the exact time dependence of biased tracers has been derived for $\Lambda$CDM and $w$CDM cosmologies, where $w$ is the equation of state parameter of a smooth dark energy component with no perturbations (see also~\cite{Fujita:2020xtd,DAmico:2021rdb}). 
In this paper, we will show how to properly account for the redshift selection. 
Then, removing both time approximations, we will analyze BOSS FS with exact time dependence. 
In Sec.~\ref{sec:beyondEdS}, after reviewing the computation of the one-loop galaxy power spectrum in redshift space with exact time dependence, we inspect the posteriors obtained by fitting simulations of volume similar to the one of DESI on $\Lambda$CDM and $w$CDM cosmologies with a BBN prior, removing the EdS approximation. 
In Sec.~\ref{sec:beyondzeff}, we focus on how to account for the redshift selection, which in principle depends on both redshifts of the galaxy pair.
We discuss how to evaluate the time dependence of the power spectrum at equal and unequal time. 
We then present the results fitting BOSS FS with redshift selection and exact time dependence.
We conclude in Sec.~\ref{sec:conclusion}. 
More details can be found in the appendices. 
We provide expressions for the exact time functions and galaxy kernels entering the one-loop expressions in App.~\ref{app:bias}. 
The expectation values of the two-point function estimators of the power spectrum and the correlation function are carefully derived in App.~\ref{app:estimator}, highlighting the redshift dependence of the observables. 
The IR-resummation at unequal time is derived in App.~\ref{app:resum}.
Finally, descriptions of the likelihood and priors used for the analyses are given in~App.~\ref{app:eftparam}. 

\begin{figure}[ht!]
\centering
\includegraphics[width=0.99\textwidth]{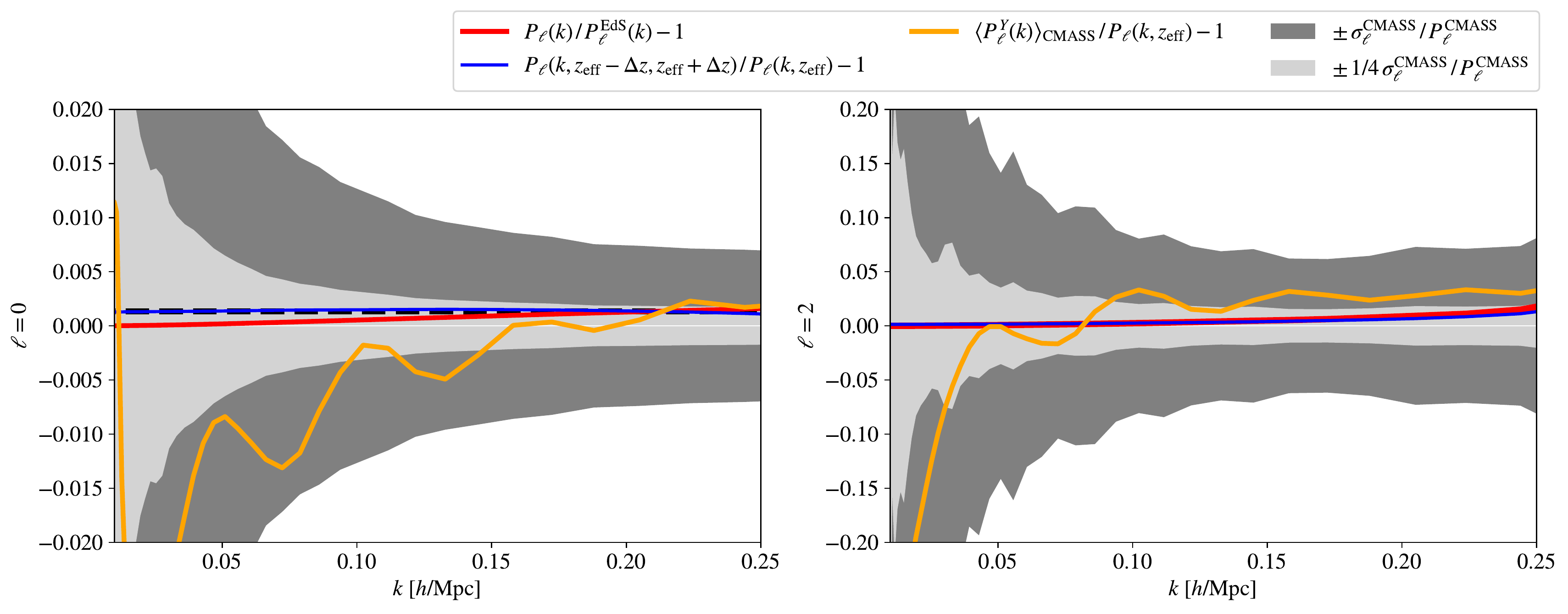}
\includegraphics[width=0.99\textwidth]{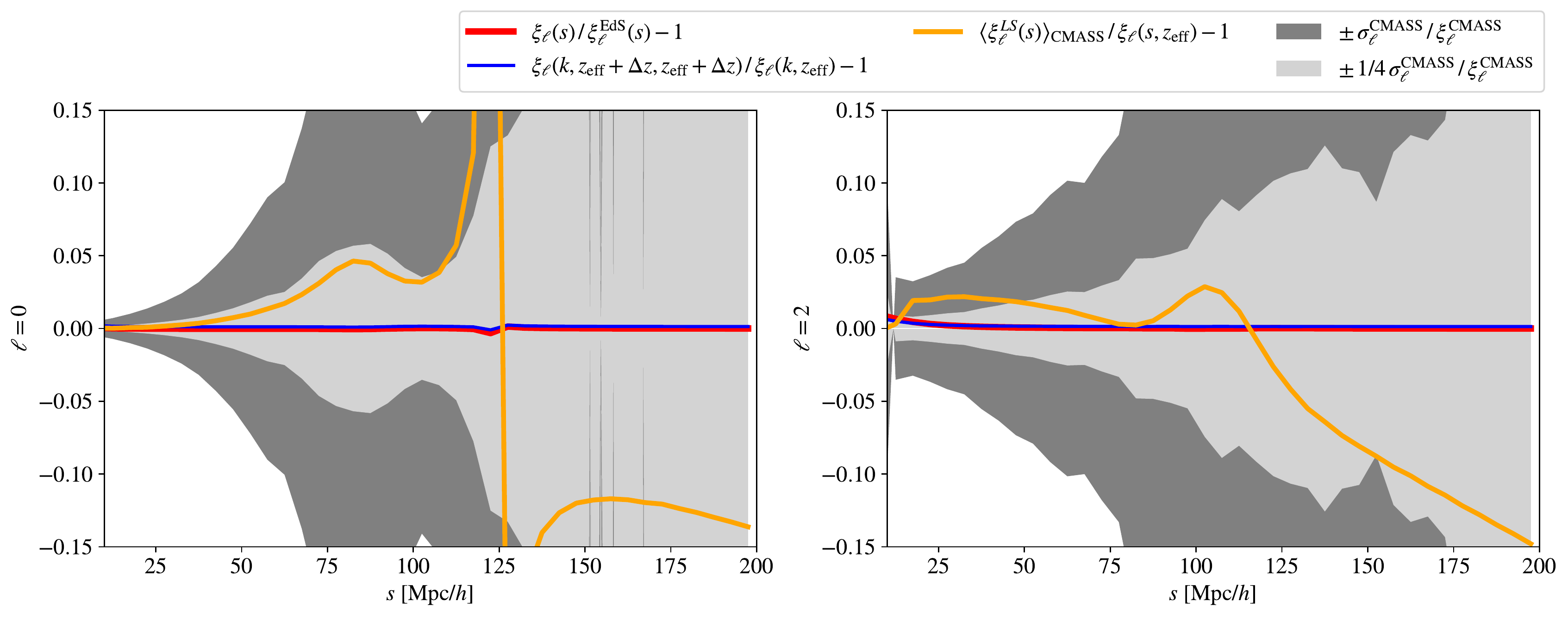}
\caption{ \footnotesize
Relative errors on the monopole and quadrupole of the redshift-space galaxy power spectrum and correlation function, evaluated on their best fit to BOSS data, of the various approximations under scrutiny in this work. 
BOSS CMASS ($0.43<z<0.7$) relative error bars are shown for comparison. 
The EdS approximation is shown in red lines. Here the evaluations beyond the EdS approximation are performed using the exact time dependence in the loop given by Eq.~\eqref{eq:density}. 
The effective redshift approximation is shown in orange lines. Here the evaluations beyond the effective redshift approximation are performed using Eq.~\eqref{eq:ps_main} and Eq.~\eqref{eq:xi_main} for the PS and the CF, respectively. 
For reference, the blue lines represent the relative differences between the two-point functions at unequal time $z_1 = z_{\rm eff} - \Delta z$ and $z_2 = z_{\rm eff} + \Delta z$, entering in the computation beyond the effective redshift approximation (and beyond the plane-parallel limit), and at equal time $z_{\rm eff}$, where $z_{\rm eff}=0.57$ is CMASS effective redshift, and $\Delta z = 0.1$ is roughly the standard deviation of CMASS redshift selection function.
}
\label{fig:pkcf}
\end{figure}

\paragraph{Approximation error summary}
For the convenience of the reader, we summarize below our main findings. 
While the difference between the EdS approximation and the exact time dependence of the one-loop galaxy two-point function in redshift space is find to be small for BOSS, the effective redshift approximation leads to an error comparable to the BOSS error bars. 
This is shown in Fig.~\ref{fig:pkcf}, where we plot the relative errors on the PS and CF multipoles associated to these approximations, and the BOSS CMASS relative error bars for comparison. 
We can see that the EdS approximation (in red lines) is rather inconsequential for BOSS. 
However, for data volume $\sim 16$ times bigger than the volume of BOSS, corresponding to the expected volume of future surveys like DESI, the error bars are decreased by 4.
In this case, we can see that the error associated to the EdS approximation becomes comparable to such error bars at $k \gtrsim 0.2 h/\textrm{Mpc}$, where the signal-to-noise ratio is the strongest. 
This therefore motivates us to check on simulation data with such large volume the accuracy of the EdS approximation, at the level of the cosmological parameters. 
We will however find that their posteriors are not shifted significantly when going beyond the EdS approximation even for such large data volume. 
In contrast, the effective redshift approximation (in orange lines) leads to, although not so big, appreciable errors, even for BOSS volume. 
We can see that the difference when going beyond the effective redshift approximation is dominated by the `masking' from the redshift selection function, rather than differences between equal or unequal time correlations (shown in blue lines). 
In particular, the masking of large separations in the direction of the line of sight impacts the BAO, which is better seen in Fourier space from the wiggles, than in configuration space where the relative error around the BAO peak is blurred by the zero-crossing of the correlation function monopole. 
Given those appreciable differences, we will check how the posteriors of the cosmological parameters obtained fitting BOSS data are shifted when going beyond the effective redshift approximation. 
We will find small shifts of $\lesssim 0.2\sigma$ in all cosmological parameters except a marginal one of about $0.4\sigma$ on $\ln(10^{10}A_s)$.

\paragraph{Data Sets} 
We make use of the SDSS-III BOSS DR12 galaxy sample~\cite{BOSS:2016wmc}. 
The BOSS CF and PS are measured from BOSS catalogs DR12 (v5) combined CMASS-LOWZ~\cite{Reid:2015gra}~\footnote{publicly available at \href{https://data.sdss.org/sas/dr12/boss/lss/}{https://data.sdss.org/sas/dr12/boss/lss/}}. 
The covariances are estimated using the patchy mocks~\cite{Kitaura:2015uqa}. 
The `lettered' challenge simulations used in this work are presented in~\cite{BOSS:2016wmc}. 
The BOSS CF and the PS, as well as their covariances, and the window functions for the PS, are the ones measured and described in~\cite{Zhang:2021yna}. 
For all the analyses in this work, we fit the monopole and quadrupole, up to $\kmax = 0.23 \hinvMpc$ when analyzing the PS, and down to $s_{\rm min} = 20 \Mpcinvh$ when analyzing the CF. 
These scale cuts were determined in~\cite{DAmico:2019fhj,Colas:2019ret,DAmico:2020kxu,Zhang:2021yna}. 
We always use a Gaussian prior on the baryon abundance $\omega_b$ centered on $0.02233$ of width $0.00036$ from BBN constraints~\cite{Mossa:2020gjc}. 

\paragraph{Public Codes}  The predictions for the FS of the galaxy power spectrum in the EFTofLSS are obtained using \code{PyBird}: Python code for Biased tracers in Redshift space~\cite{DAmico:2020kxu}~\footnote{\href{https://github.com/pierrexyz/pybird}{https://github.com/pierrexyz/pybird}}. 
The exact time dependence is made available in \code{PyBird}. 
The linear power spectra are computed with the \code{CLASS} Boltzmann code~\cite{Blas_2011}~\footnote{ \href{http://class-code.net}{http://class-code.net}}.
The posteriors are sampled using the \code{MontePython} cosmological parameter inference code~\cite{Brinckmann:2018cvx, Audren:2012wb}~\footnote{ \href{https://github.com/brinckmann/montepython\_public}{https://github.com/brinckmann/montepython\_public}}.
The triangle plots are obtained using the \code{GetDist} package~\cite{Lewis:2019xzd}. 
The FS of BOSS CF and PS, as well as the ones from the patchy mocks for the covariance, were measured~\cite{Zhang:2021yna} using \code{FCFC} and \code{powspec}, respectively~\cite{Zhao:2020bib}~\footnote{\href{https://github.com/cheng-zhao/FCFC}{https://github.com/cheng-zhao/FCFC} ; \href{https://github.com/cheng-zhao/powspec}{https://github.com/cheng-zhao/powspec}}. 
The PS window functions were measured in~\cite{Zhang:2021yna} as described in~\cite{Beutler:2018vpe} using \code{nbodykit}~\cite{Hand:2017pqn}~\footnote{\href{https://github.com/bccp/nbodykit}{https://github.com/bccp/nbodykit}}.

\section{Beyond the EdS time approximation}\label{sec:beyondEdS}

In this section, after reviewing the computation of the galaxy power spectrum at one-loop order in redshift space with exact time dependence, we check the difference introduced by the EdS time approximation at the level of the cosmological parameter posteriors by fitting simulations of volume about $50 \, {\rm Gpc}^3$.

\subsection{Biased tracers in redshift space with exact time dependence}

Biased tracers in redshift space (e.g. observed galaxies) within the EFTofLSS have previously been described in~\cite{Perko:2016puo} and with exact time dependence in~\cite{Donath:2020abv}. 
Under the EdS approximation, all time dependence of biased tracers can be entirely captured by a set of EFT parameters, that are free, time-dependent coefficients, as well as powers of the growth factor $D$ and its first log-derivative, the growth rate $f = \frac{\mathrm{d} \ln D}{\mathrm{d} \ln a}$.
This no longer holds when working with exact time dependence.
\paragraph{EFT expansion} Removing the EdS approximation, in the EFT expansion~\footnote{By `EFT' expansion, we conglomerate in one name an expansion in several parameters, such as the size of matter overdensities, the size of the short distance displacements, the derivative expansion in the size of the galaxies, etc., which is ordinarily done when solving perturbatively the EFTofLSS equations.} (up to third order) of the galaxy density field $\delta_g$, an additional operator $\mathbb{C}^{(3)}_{Y}$ appears at third order~\cite{Donath:2020abv}:
\begin{equation}\label{eq:density}
\delta_g (\k, a) = \delta_g^{\rm EdS}(\k, a) + b_1 Y(a) \mathbb{C}^{(3)}_{Y}(\k),
\end{equation}
where $\delta_g^{\rm EdS}$ is the galaxy EFT expansion under the EdS approximation and $a$ is the scale factor. 
The time dependence of the new operator is proportional to $b_1$, the linear galaxy bias, multiplied by a calculable function of time, $Y$, that vanishes in the EdS approximation, thus adding no new EFT parameter. 
As for the galaxy velocity divergence,~$\theta_g$, since up to higher-derivative terms there is no velocity bias, we can describe the galaxy velocity divergence as a specific species of biased tracer with calculable coefficients, which also includes the additional operator $C_Y$ multiplied by $Y$. Explicit expressions for $\delta_g$, $\theta_g$, and the time functions entering in their expansion can be found in App.~\ref{app:bias}. 

From there, one follows the standard steps to obtain the one-loop galaxy power spectrum in redshift space.
By switching coordinates from real space to redshift space, the galaxy {EFT} expansion in redshift space mixes real-space density and velocity operators.
At each order $n$ in perturbations, the redshift-space galaxy density field can be written as:
\begin{equation}
\delta_{g,r}^{(n)}(\k, a) = \int \, \frac{d^3 q_1}{(2\pi)^3} \dots \frac{d^3 q_n}{(2\pi)^3} \, \delta_D(\k -\q_1 - \dots \q_n) Z_n(\q_1, \dots, \q_n, a) \delta_{\q_1}^{(1)}(a) \dots \delta_{\q_n}^{(1)}(a),
\end{equation}
where $\delta_D$ is the delta Dirac function and $\delta_{\q}^{(1)}$ is the linear matter density field. 
Dropping the time variable in the notation when it is clear from the context, the redshift-space galaxy density kernels $Z_1,Z_2$, and $Z_3$, without the counterterms, read~\cite{Perko:2016puo}:
\begin{align}\label{eq:redshift_kernels}\nonumber
    Z_1(\q_1) & = K_1(\q_1) +f\mu_1^2 G_1(\q_1) = b_1 + f\mu_1^2,\\ \nonumber
    Z_2(\q_1,\q_2,\mu) & = K_2(\q_1,\q_2) +f\mu_{12}^2 G_2(\q_1,\q_2)+ \, \frac{1}{2}f \mu q \left( \frac{\mu_2}{q_2}G_1(\q_2) Z_1(\q_1) + \text{perm.} \right),\\ \nonumber
    Z_3(\q_1,\q_2,\q_3,\mu) & = K_3(\q_1,\q_2,\q_3) + f\mu_{123}^2 G_3(\q_1,\q_2,\q_3) \nonumber \\ 
    &+ \frac{1}{3}f\mu q \left(\frac{\mu_3}{q_3} G_1(\q_3) Z_2(\q_1,\q_2,\mu_{123}) +\frac{\mu_{23}}{q_{23}}G_2(\q_2,\q_3)Z_1(\q_1)+ \text{cyc.}\right),
\end{align}
where $\mu= \q \cdot \hat{\z}/q$, $\q = \q_1 + \dots +\q_n$, and $\mu_{i_1\ldots  i_n} = \q_{i_1\ldots  i_n} \cdot \hat{\z}/q_{i_1\ldots  i_n}$, $\q_{i_1 \dots i_m}=\q_{i_1} + \dots +\q_{i_m}$, with $\hat{\z}$ being the unit vector in the direction of the line of sight, and $n$ is the order of the kernel $Z_n$. 
Here $K_n$ and $G_n$ are the galaxy density and velocity kernels, respectively, that are given in App.~\ref{app:bias}.  
All kernels entering the power spectrum at one loop can be described with 4 galaxy bias parameters~$\lbrace b_i \rbrace$ after UV-subtraction~\cite{Senatore:2014eva,Angulo:2015eqa,Fujita:2016dne}. 

We find useful to highlight the exact time dependence at this stage. 
Removing the EdS approximation, the galaxy density kernels are modified at third order to (see details in App.~\ref{app:bias}):
\begin{align}
K_3(\q_1, \q_2, \q_3, a, \lbrace b_i \rbrace) & = K_3^{\rm EdS}(\q_1, \q_2, \q_3 , \lbrace b_i \rbrace) + b_1 Y(a) K_Y(\q_1, \q_2, \q_3).
\end{align}
Furthermore, the velocity divergence can be written as a biased density tracer with specific values for the bias coefficients~$\lbrace \tilde b_i \rbrace$~\cite{Perko:2016puo}, yielding:
\begin{equation}\label{eq:G_n}
G_n(\q_1, \dots, \q_n, a) = K_n(\q_1, \dots, \q_n, a, \lbrace \tilde b_i \rbrace).
\end{equation}
In the basis of descendants~\cite{Angulo:2015eqa,Fujita:2016dne}, we have~\cite{Donath:2020abv}:
\begin{equation}\label{eq:replacement}
\tilde b_1 = 1, \quad \tilde b_2 = -\frac{7}{2} \mG_1^\theta(a), \quad \tilde b_3 = 21 \, \mV_{12}^{\theta}(a), \quad \tilde b_4 = -\frac{5}{2} + \frac{7}{2} \mG_1^\theta(a),
\end{equation}
where the time functions $\mG_1^\theta(a), \mV_{12}^{\theta}(a)$ stem from the exact solutions to the equations of motion of the dark matter fields, and are given in Eq.~\eqref{eq:timefunc}. 

\paragraph{Galaxy power spectrum} Adding the counterterms and the stochastic terms, the one-loop galaxy power spectrum in redshift space is then given by~\cite{Perko:2016puo}:
\begin{align}\label{eq:powerspectrum}
& P(k, \mu) = Z_1(\mu)^2 P_{11}(k) \nonumber  \\
& + 2 \int \frac{d^3q}{(2\pi)^3}\; Z_2(\q,\k-\q,\mu)^2 P_{11}(|\k-\q|)P_{11}(q) + 6 Z_1(\mu) P_{11}(k) \int\, \frac{d^3 q}{(2\pi)^3}\; Z_3(\q,-\q,\k,\mu) P_{11}(q)\nonumber \\
& + 2 Z_1(\mu) P_{11}(k)\left( c_\text{ct}\frac{k^2}{{ k^2_\textsc{m}}} + c_{r,1}\mu^2 \frac{k^2}{k^2_\textsc{m}} + c_{r,2}\mu^4 \frac{k^2}{k^2_\textsc{m}} \right) + \frac{1}{\bar{n}_g}\left( c_{\epsilon,0}+c_{\epsilon,1}\frac{k^2}{k_\textsc{m}^2} + c_{\epsilon,2} f\mu^2 \frac{k^2}{k_\textsc{m}^2} \right) \, ,
\end{align}
where $P_{11}$ is the linear matter power spectrum, $\mu$ is the cosine of the angle of the wavenumber $\k$ with the line of sight, $k^{-1}_\textsc{m} (\simeq k^{-1}_\textsc{nl})$ is the scale controlling the bias (dark matter) derivative expansion, and $\bar{n}_g$ is the mean galaxy number density.
In the first line it appears the linear contribution. 
The one loop contribution is given in the second line. 
In the third line, the first terms are the counterterms: the term in $c_\text{ct}$ is a linear combination of the dark matter speed of sound~\cite{Baumann:2010tm,Carrasco:2012cv} and a higher derivative bias~\cite{Senatore:2014eva}, and the terms in $c_{r,1}$ and $c_{r,2}$ represent the redshift-space counterterms~\cite{Senatore:2014vja}. 
The last terms are the stochastic contributions~\cite{Perko:2016puo}. 

\paragraph{IR-resummation} Finally, non-perturbative bulk displacements need to be resummed~\cite{Senatore:2014via}. 
In App.~\ref{app:resum}, we derive the IR-resummation for unequal-time correlation, from which the following formula can be read off. 
At equal time, the IR-resummed $N$-loop power spectrum in redshift space in terms of its multipole moments $P_\ell(k)|_{N}$ is given by~\cite{Senatore:2014vja,Lewandowski:2015ziq}:
\begin{align}
P_\ell(k)|_{N} & = \sum_{j=0}^N \sum_{\ell'}  4\pi (-i)^{\ell'} \int dr \, r^ 2 \, Q_{||N-j}^{\ell \ell'}(k,r) \, \xi_{\ell'}^{j} (r), \label{eq:resumConvol}\\
\xi_{\ell'}^{j} (r) & = i^{\ell'}  \int  \frac{dp\, p^2}{2\pi^2} P_{\ell'}^{j}(p)  \, j_{\ell'}(pr),
\end{align}
where $P_{\ell}^{j}$ and $\xi_{\ell}^{j}$ are the $j$-loop order pieces of the Eulerian ({\it i.e.} non-resummed) power spectrum and correlation function, respectively. 
The effects from the bulk displacements are encoded in: $Q_{||N-j}^{\ell \ell'}(k,r) \sim e^{-k^2 A_0(r)/2}$ (see App.~\ref{app:resum} for explicit expressions).
Expanding $Q_{||N-j}^{\ell \ell'}(k,r)$ in powers of $k^2$, the resummed power spectrum $P_\ell(k)|_N$ can be written as a sum of the non-resummed power spectrum $P_\ell(k) $ plus `IR-corrections'~\cite{DAmico:2020kxu}: 
\begin{equation}\label{eq:resum}
P_\ell(k)|_N = P_\ell(k) + \sum_{j=0}^N \sum_{\ell'} \sum_{n=1} \sum_{\alpha} 4\pi (-i)^{\ell'} k^{2n} \,\mathcal{Q}_{||N-j}^{\ell \ell'}(n, \alpha) \, \int dr \, r^ 2 \,  \left[ A_i(r) \right]^n   \xi^j_{\ell'}(r) \, j_{\alpha}(kr) \, ,
\end{equation}
where $n$ is controlling the expansion of $Q_{||N-j}^{\ell \ell'}(k,r)$ in powers of $k^2$, $\alpha$ is the order of the spherical Bessel function $j_\alpha$ running over $\lbrace 0, 2, 4, ... \rbrace$, $\mathcal{Q}_{||N-j}^{\ell \ell'}(n, \alpha)$ is a number depending on $N-j$, $\ell$, $\ell'$, $n$ and $\alpha$ (and is function of $f$), and $\left[ A_i(r) \right]^n$ denotes a product of the form $A_0(r) \times ...  \times A_0(r) \times A_2(r) \times ... \times A_2(r)$ such that the total number of terms in the product is $n$.

\paragraph{} The difference between the exact time computation and the EdS one is shown in Fig.~\ref{fig:pkcf}.  
At $k = 0.23 \hinvMpc$, it is a few permils for the monopole and around $1\%$ for the quadrupole, as shown in~\cite{Donath:2020abv}. 
Compared to CMASS error bars, the difference is relatively small, about $\lesssim 1/4 \sigma_{\rm data}$ at $k = 0.23 \hinvMpc$ in both the monopole and the quadrupole. 
We thus expect to find no difference at the level of the posteriors in the fit to BOSS data. 
However, we now check on simulations of volume about $12$ times the one of CMASS, for which the difference between the exact time computation and the EdS one becomes comparable to the error bars. 

\begin{figure}
\centering

\includegraphics[width=0.99\textwidth]{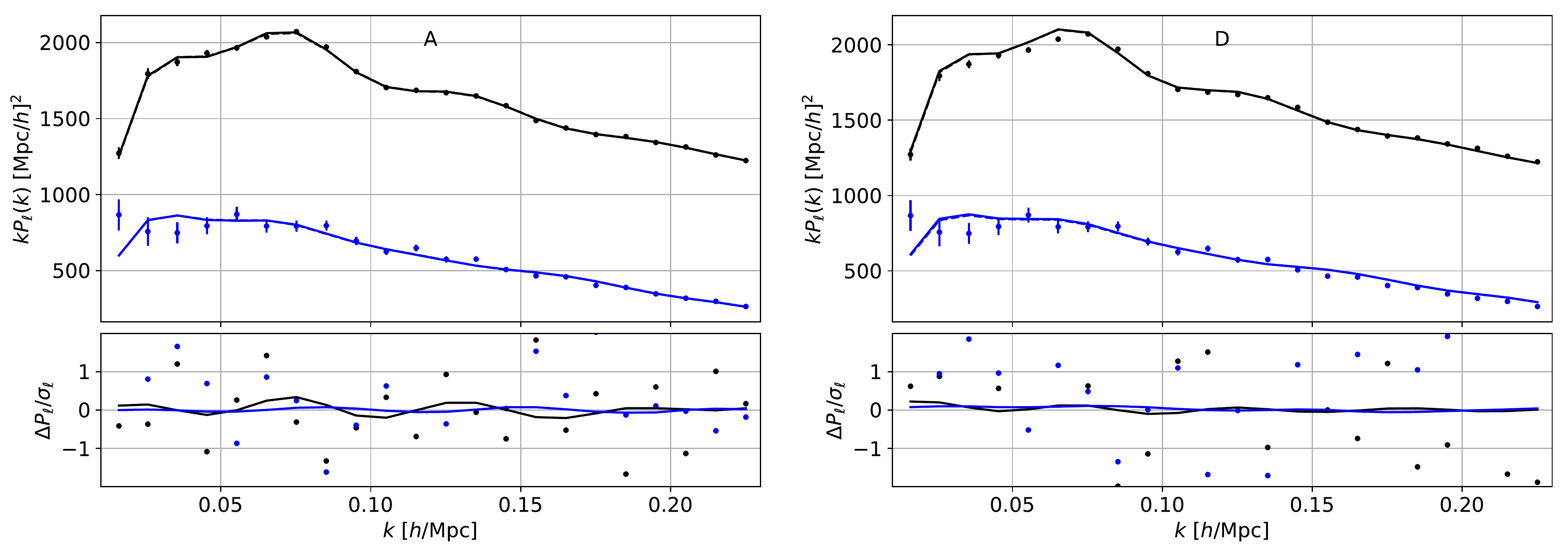}
\caption{\small Best fits and residuals obtained with exact time dependence on the power spectrum multipoles of the lettered challenge simulations A and D on $\Lambda$CDM with a BBN prior. 
The monopole is displayed in black and the quadrupole in blue.
The data error bars depicted are the {square root of the} covariance diagonal elements.
In the lower panels, the relative difference with the best fit obtained with the EdS approximation is shown in continuous lines.
The oscillations displayed here stem from the difference in the best fit values, in particular in $\Omega_m$ that controls the BAO angle (see Fig.~\ref{fig:challenge}), and not from the difference between the exact-time and EdS evaluation, that is rather smooth, as displayed in Fig.~\ref{fig:pkcf} for fixed cosmological and EFT parameters. }
\label{fig:challenge_bestfit}
\end{figure}

\subsection{Tests on simulations}
\begin{figure}[h]
\centering
\small
\includegraphics[width=0.49\textwidth]{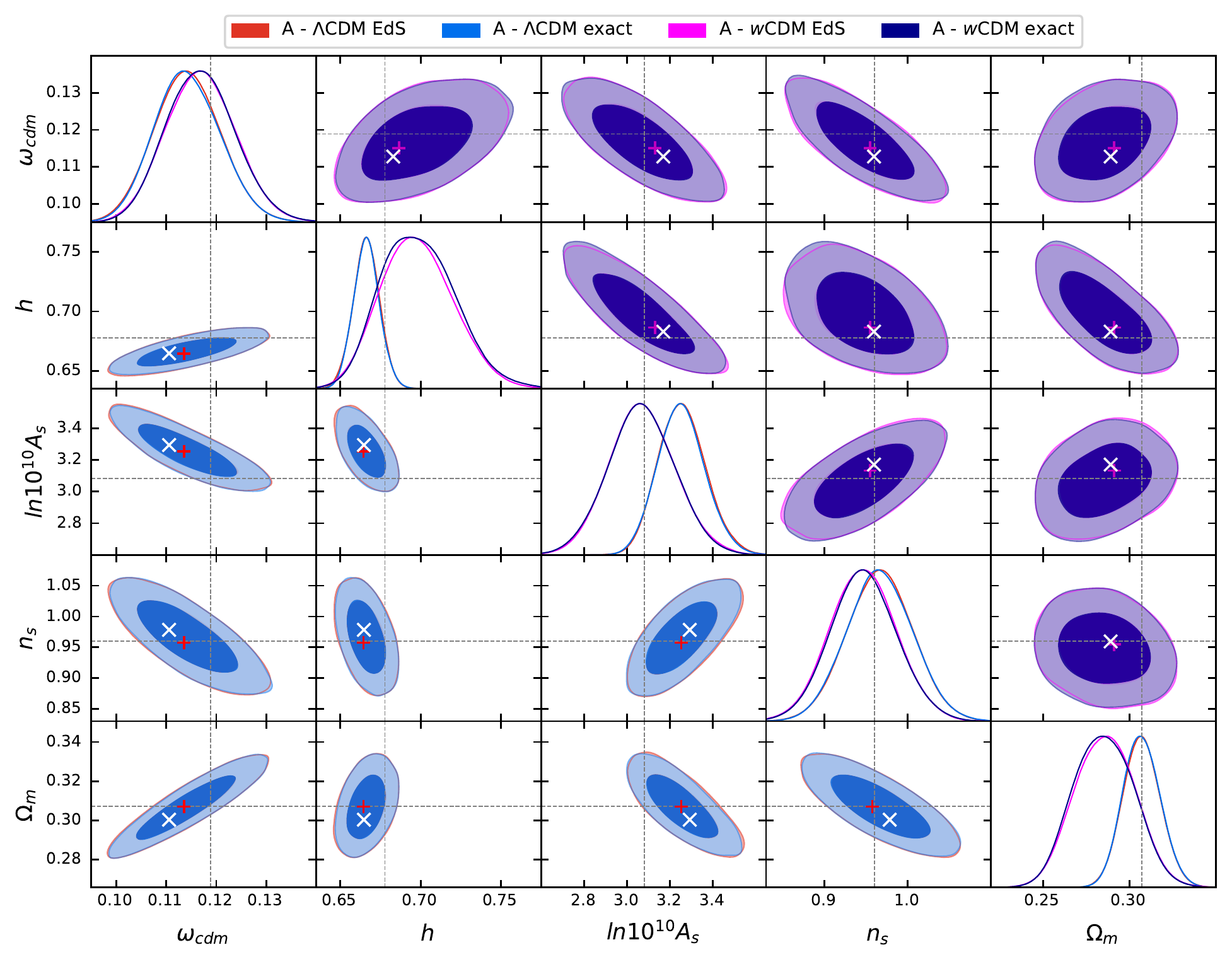}
\includegraphics[width=0.49\textwidth]{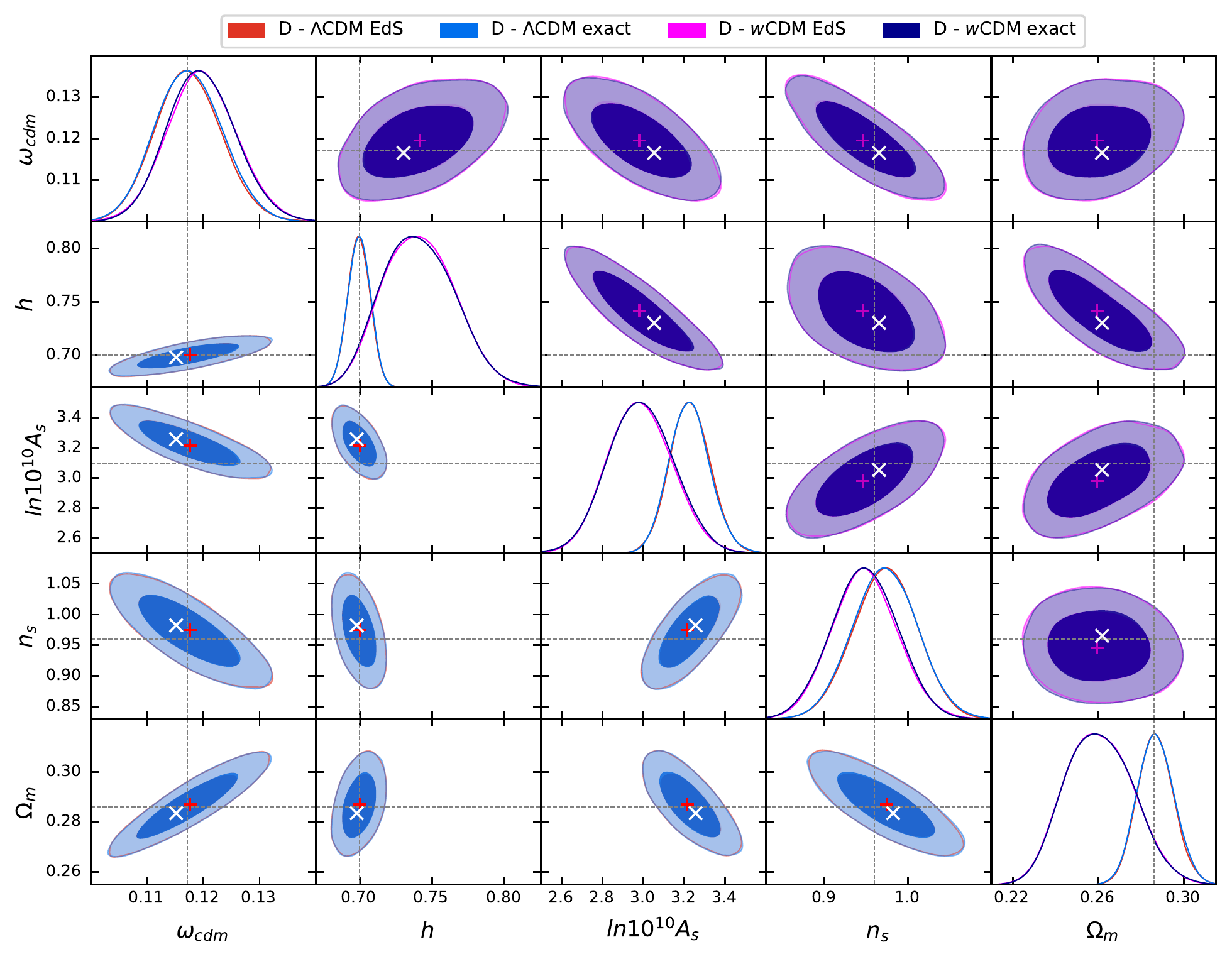}
\includegraphics[width=0.49\textwidth]{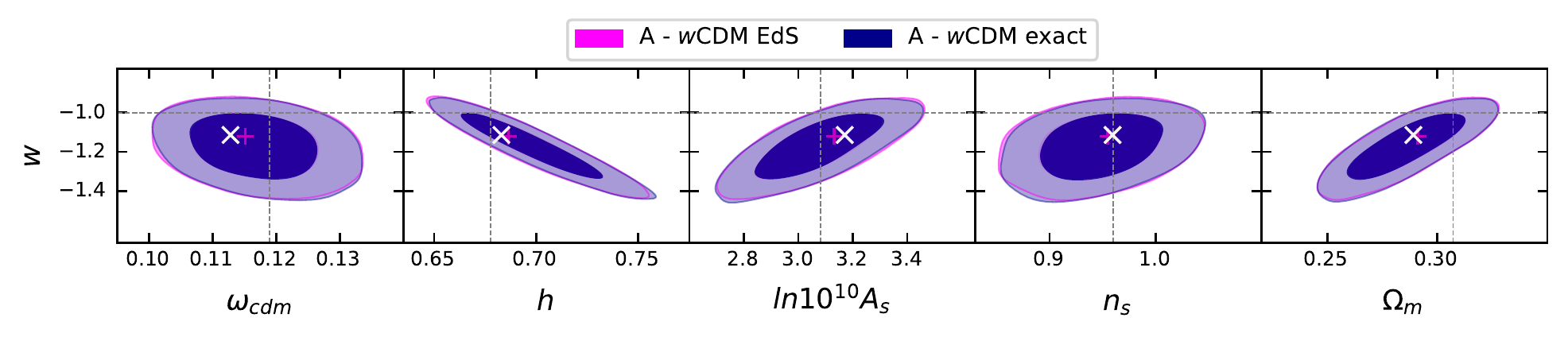}
\includegraphics[width=0.49\textwidth]{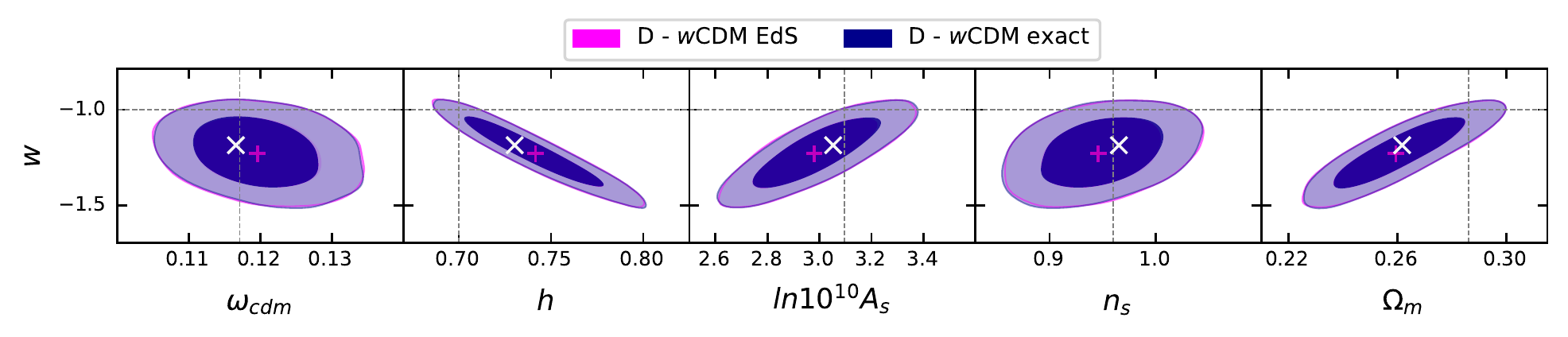}
\caption{Posteriors of the cosmological parameters obtained fitting the power spectrum multipoles of the lettered challenge simulations A and D on $\Lambda$CDM and $w$CDM with a BBN prior, with and without the EdS approximation. 
The best fit values are depicted as `$+$' and `$\times$', respectively.  
} 
\label{fig:challenge}
\end{figure}

Several sets of simulations were analyzed to measure the performances of the EFTofLSS, see e.g.~\cite{DAmico:2019fhj,Colas:2019ret,DAmico:2020kxu,Nishimichi:2020tvu}.
Here, we make use of the BOSS `lettered challenge' N-body simulations, that were part of the partly-blind BOSS challenge presented in~\cite{BOSS:2016wmc}. 
These were used to measure the theory-systematic errors on BOSS data in~\cite{DAmico:2019fhj,Colas:2019ret,DAmico:2020kxu,Zhang:2021yna}. Those are periodic boxes of side length $2.5 {\rm Gpc}/h$, populated with various high-fidelity halo occupation distributions, and are described in e.g.~\cite{DAmico:2019fhj}. 
The theory-systematic errors were found to be at most $\lesssim 1/3 \sigma_{\rm data}$ for all cosmological parameters at $\kmax = 0.23 \hinvMpc$ both for $\Lambda$CDM and $w$CDM with a BBN prior in~\cite{Colas:2019ret} and~\cite{DAmico:2020kxu}, respectively, where $\sigma_{\rm data}$ are the error bars obtained fitting BOSS data. 
In this work, we compare the posteriors obtained fitting the lettered challenge simulations A and D on $\Lambda$CDM and $w$CDM, removing the EdS time approximation. 
The redshifts of A and D are $z=0.562$ and $z=0.5$, respectively. 
The likelihood, priors and EFT parameters are presented in App.~\ref{app:eftparam}. 
The best fits are shown in Fig.~\ref{fig:challenge_bestfit}.
The triangle plots of the cosmological parameters are given in Fig.~\ref{fig:challenge}. 

\paragraph{} The difference in the best fit at the level of the power spectrum is of the same order as the shifts in the posteriors, which is relatively small in terms of the error bars. 
Although marginal, there is a slight improvement of the minimum $\chi^2$, about $0.4$ for A and $2.0$ for D. 
The difference in the 68\% and the 95\% contours is however barely visible. 
This means that it is safe to use the EdS time approximation for data of volume as big as $\sim 50 {\rm Gpc}^3$. 
For BOSS, which has a much smaller volume, it is trivial to check that the same conclusion holds.  Overall, as the change of the posteriors are negligible, we find that the systematic error is negligible for the BOSS analysis, as already concluded from the studies in the EdS approximation in~\cite{{DAmico:2019fhj,Colas:2019ret,DAmico:2020kxu}}.

\paragraph{} Before moving to the next section, we finish with a comment on $w$CDM. 
Here $w$CDM refers to smooth dark energy, that is, with no perturbations.
This means that the speed of sound of dark energy fluctuations, $c_s^2$, goes to one, $c_s^2 \rightarrow 1$, such that the sound horizon $\mathcal{H}/c_s$ is about the size of the cosmological horizon. 
 The effective field theory of dark energy~\cite{Creminelli:2006xe} shows that, as $c_s^2 \rightarrow 1$, $w<-1$ corresponds to an unstable vacuum with instable perturbations (see also{~\cite{Cline:2003gs} for early statements that a theory with a ghost is not viable}). 
From a Bayesian point of view, one should therefore impose a prior on $w$ to restrict it to the physically allowed region. 
In the current work, as our main focus is to highlight the exact-time dependence, we show for illustration in Fig.~\ref{fig:challenge} the posteriors with no prior on $w$.  
The analysis of the cosmological data with $w < -1$ on $w$CDM or with a consistent treatment of the gravitational potential in the presence of dark energy fluctuations in the limit $c_s^2 \rightarrow 0$, i.e. clustering quintessence is presented in~\cite{DAmico:2020tty}.

\section{Beyond the effective redshift approximation}\label{sec:beyondzeff}
\begin{figure}[h]
\centering

\includegraphics[width=0.5\textwidth]{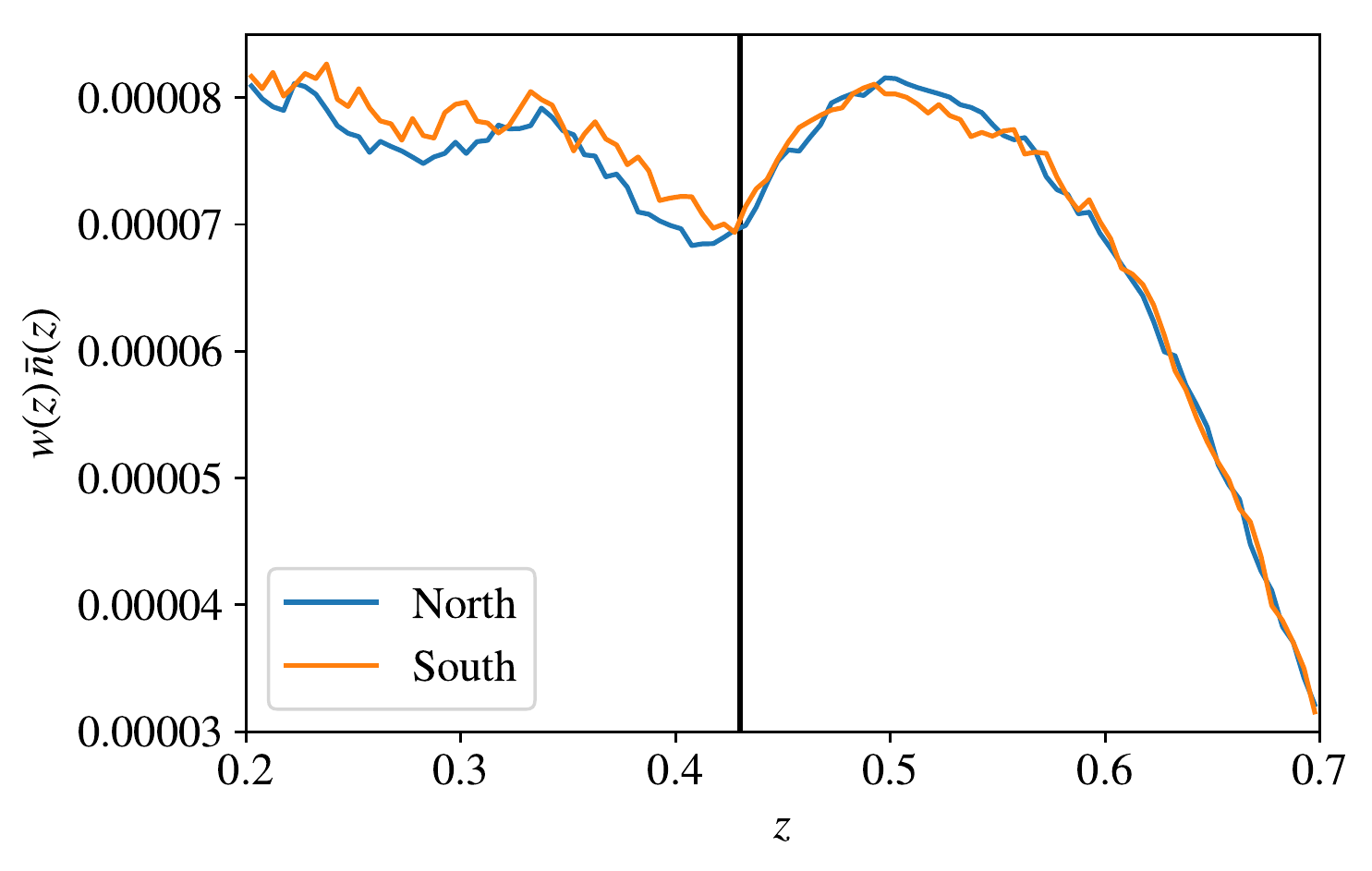}
\caption{ \small
BOSS redshift selection functions, i.e. galaxy number counts distribution $\bar n(z)$ times correction weights $w(z)$ (mainly FKP), up to a normalization factor. 
The vertical line is at $z=0.43$, where we choose to cut the data into two samples, $0.2 < z < 0.43$, and $0.43 < z < 0.7$, referred as LOWZ and CMASS, respectively. 
Here we have used the reference cosmology $\Omega_m^{\rm ref} = 0.310$ for the coordinates-to-distance conversion, and $P_0= 10^{4} \, (\textrm{Mpc}/h)^3$ to compute the FKP weights. 
}
\label{fig:nz}
\end{figure}

Galaxy surveys data are gathered into redshift bins.
Over the selected redshift bins, the distribution of galaxy number counts varies. This is illustrated for the BOSS samples LOWZ and CMASS, respectively $0.2<z<0.43$ and $0.43<z<0.7$, in Fig.~\ref{fig:nz}. 

While all redshifts of the galaxies in spectroscopic surveys are measured to great precision, usually, the theory prediction is evaluated at a single effective redshift, to compare with the data in a given redshift bin (see~e.g.~\cite{DAmico:2019fhj}). 
Assuming a single effective redshift allows for fast evaluation, however, is obviously an approximation, as it boils down to assume that all the galaxies are at the same redshift: this is not the case, as seen in Fig.~\ref{fig:nz}. 
In this section, we check the accuracy of this approximation, by eliminating this approximation. 

The power spectrum in redshift space is usually measured using the Yamamoto estimator~\cite{Yamamoto:2005dz}. 
To perform the measurements in an efficient manner using fast Fourier transforms, the line of sight can be chosen as the direction of one of the galaxy in each pair~\cite{Bianchi:2015oia,Scoccimarro:2015bla}.
The expectation value then reads:
\begin{equation}\label{eq:ps_main}
\braket{P_\ell^Y(k)} = \frac{2\pi (2\ell+1)}{N_P} (-i)^\ell \int ds \, s^2 j_\ell(ks)  \int_{-1}^{+1} d\mu \, \int dr_1 \, r_1^2 \, \bar n(r_1) \bar n(r_2)  \xi(s, \mu, r_1) \, \mathcal{L}_\ell(\mu) Q(s, \mu, r_1) \, , 
\end{equation}
where $\s$ is the separation, $\mu  \equiv \hat s \cdot \hat r_1$ the cosine of its angle with the line of sight $\r_1$, $r_2 \equiv \sqrt{s^2 + r_1^2 + 2 s r_1 \mu}$, and $N_P \equiv \int d^3 r \, \bar n(\r)^2$ is a convenient normalization factor. 
$\mathcal{L}_\ell$ denotes the Legendre polynomial of order $\ell$, and $Q(s, \mu, r_1)$ is the survey geometry window function, that generically depends on the separation $s$ of the two objects in each pair, and the line-of-sight distance $r_1$ and orientation $\mu$. 
Here we have redefined $\bar n(r) \equiv w(r) \bar n(r)$ for conciseness. 

As it is computationally challenging to obtain the 3D window function, in the following we will set $Q(s, \mu, r_1) \equiv 1$ when comparing Eq.~\eqref{eq:ps_main} with the effective redshift approximation. 
When determining cosmological parameters, we will instead work in configuration space with the correlation function, where a full account of the redshift evolution is not a prohibitive task. 
Indeed, in the Landy \& Szalay estimator for the correlation function~\cite{1993ApJ...412...64L}, the effect of the mask nicely cancels out, and therefore, it does not need to be applied on the theory model. 
Choosing the line of sight to be the mean direction of the pair, $\r \equiv \frac{1}{2}(\r_1 + \r_2)$, the expectation value of the correlation function estimator reads:
\begin{equation}\label{eq:xi_main}
\braket{\xi_\ell^{\rm LS}(s) }  = \frac{2\pi (2\ell + 1)}{N_\xi(s)} \int dr \, r^2 \int_{-1}^{+1} d\mu \, \bar n(r_1) \bar n(r_2) \xi(s, \mu, r) \mathcal{L}_\ell(\mu)\,  ,
\end{equation}
where $\s$ is the separation, $\mu  \equiv \hat s \cdot \hat r$ the cosine of its angle with the line of sight $\r$, $r_{1,2} \equiv \sqrt{r^2 + (s/2)^2 \mp s r \mu}$, and $N_\xi(s) =  2\pi \int dr \, r^2 \int_{-1}^{+1} d\mu \, \bar n(r_1) \bar n(r_2)$ is the normalization factor. 
Here we have also redefined $\bar n(r) \equiv w(r) \bar n(r)$ for conciseness. 

In App.~\ref{app:estimator}, we provide a derivation of the expectation value of these two estimators used to measure the two-point function from the data, both for the correlation function and the power spectrum. 
Note that we are neglecting the so-called wide-angle corrections in the expression above, and we provide some comments in the appendix. 

The relative difference from Eq.~\eqref{eq:ps_main} for the power spectrum, and Eq.~\eqref{eq:xi_main} for the correlation function, to their evaluation at a single effective redshift $z_{\rm eff}$, is shown for BOSS CMASS in Fig.~\ref{fig:pkcf}. 
The difference is appreciable, but rather small, of $\lesssim 1/2 \sigma_{\rm CMASS}$, at most. 
Still, this shows that to use the effective redshift approximation, one needs to be careful when selecting redshift bins. 
In the following, after providing details on how we evaluate Eqs.~\eqref{eq:ps_main}~and~\eqref{eq:xi_main}, we will show the impact on the determination of the cosmological parameters from BOSS data.

\subsection{Biased tracers in redshift space at unequal time}\label{sec:redshiftselection}

The double integrals over $\mu$ and $r_1$ (or $r$) present in Eq.~\eqref{eq:ps_main} or Eq.~\eqref{eq:xi_main} can be performed by two nested trapezoidal integrations on a meshgrid of $s$, $\mu$ and $z$. 
For the power spectrum, we are further left with a simple spherical-Bessel transform that can be performed with a FFTLog. 
All this computation relies on the unequal-time correlation function $\xi(s, \mu, r)$, depending on the galaxies separation and the distance of the observer to the galaxy pair.
Let us now give explicitly the expression of the unequal-time two-point function. 
As the perturbative expansions presented in Section~\ref{sec:beyondEdS} are written in Fourier space, where they read more naturally, we will continue to write the expressions for the power spectrum, at unequal time, in the following. 
To obtain (the time-independent part of) the configuration-space correlation function, which is merely an inverse spherical-Bessel transform of (the time-independent part of) the power spectrum, we can follow the computational strategy described in~\cite{Zhang:2021yna}, allowing us to evaluate the correlation function with the same computational complexity as the evaluation of the power spectrum. 

We will write the expressions as a functions of the redshifts $z_1$ and $z_2$ of the two galaxies in a pair. 
These are functions of the comoving distances, $z_1(r_1)$ and $z_2(r_2)$. The distances of the galaxies to the observer, $r_1$ and $r_2$, need to satisfy the condition that the observer and the two galaxies form a triangle, and depend on the choice of the line of sight. 
Explicitly, the conversion is as follow. 
For the power spectrum estimator, we are using the end-point line of sight $\r_1$, such that $r_1 = \chi(z)$ and $r_2 \equiv \sqrt{s^2 + r_1^2 + 2 s r_1 \mu}$, where $\chi(z)$ is the comoving distance as function of redshift. 
For the correlation function estimator, we are using the mean line of sight $\r$, such that $r_{1,2} \equiv \sqrt{r^2 + (s/2)^2 \mp s r \mu}$, where $r = \chi(z)$. 
See more details in App.~\ref{app:estimator}.

\paragraph{Equal time} Before moving to the unequal-time case, it is first instructive to look a the equal-time correlator. 
As long as all degrees of freedom of the universe have a negligible speed of sound or mean free path, the time dependence of the power spectrum factorizes. 
In the EdS approximation, the galaxy power spectrum, Eq.~\eqref{eq:powerspectrum}, can be written as a sum of products of time functions multiplied by time-independent pieces $P_\eta(\k)$:
\begin{equation}\label{eq:facto}
P(\k, z) = \sum_\eta b_i^{\mu_\eta} b_j^{\nu_\eta} D(z)^{2\rho_\eta} f(z)^{\sigma_\eta} P_\eta(k,\mu) \, ,
\end{equation}
where $\{ \mu_\eta, \nu_\eta \} \in \{ 0, 1 \}$ (biasing), $\{ \rho_\eta \} \in \{ 1, 2 \}$ (loop order), $\{ \sigma_\eta \} \in \{ 0, 1, 2, 3, 4 \}$ (redshift space distortions).
Here $b_i, b_j$ represent EFT parameters, $D$ is the growth factor and $f$ is the growth rate. 
For example, the linear contribution is:
\begin{align}
P_{\rm lin}(\k, z) & = (b_1 + f (z)\mu^2)^2 P_{11}(k,z) \nonumber \\
 & = b_1^2 D(z)^{2} P_{\rm lin,0}(k) + 2 b_1 f(z) D(z)^{2} \mu^2 P_{\rm lin,0}(k) + f(z)^2 D(z)^{2} \mu^4 P_{\rm lin,0}(k) \, ,
\end{align}
where $P_{\rm lin,0}(k)$ is the linear matter power spectrum at $z=0$, such that $D(0) \equiv 1$. 
The time dependence of the IR resummation, once put in the form of Eq.~\eqref{eq:resum}, can be factorized in the same way.
In particular, the time dependence of the IR corrections is:
\begin{equation}
\int dr \, r^ 2 \, \left[ A_i(r,z) \right]^n   \xi^j_{\ell'}(r,z)  \, j_{\alpha}(kr) = D(z)^{2n+2(j+1)} \int dr \, r^ 2 \,  \left[ A_{i,0}(q) \right]^n \xi^{j}_{\ell',0}(r)  \, j_{\alpha}(kr) \, ,
\end{equation}
where a quantity with a subscript `$0$' is evaluated at $z=0$. 

Importantly, here and after, we are assuming that the time dependence of the EFT parameters is mild within the selected redshift bin, such that they can be taken out of the integration. 
Indeed, the time dependence of the EFT parameters can be related to dark matter halo and galaxy formation physics. 
Typically, the linear galaxy bias $b_1$ is found to evolve as $D(z)^{-p}$, where $p$ is a number of $\mathcal{O}(1)$, see e.g.~\cite{Papageorgiou:2012jg}. 
This allows us to estimate the variation of $b_1$ along CMASS or LOWZ bin: roughly, around their respective effective redshift, to $[z_{\rm min}, z_{\rm max}] = [0.2, 0.43]$ or $[0.43, 0.7]$, respectively, $b_1$ varies by about $\pm 8\%$. 
This is already smaller in size to the relative $68\%$-confidence intervals that we will obtain on $b_1$ fitting BOSS data, which are about $\pm 10\%$ (see Table~\ref{tab:boss}). 
Given that the effective value of $b_1$ will be an average of the values it takes along the bin, with most weight around the effective redshift, the mistake associated to the assumption that, $b_1$, and similarly the other EFT parameters, are constant along the redshift bins, is expected to be relatively small compared to the error bars we will obtain. 
As the time evolution of the galaxies biases involves small-scales physics beyond the EFT paradigm (and thus beyond the scope of this paper), we leave precise quantification of their effect on the observables and the cosmological parameters to future work. 

Let us now add another comment on the time evolution of the two-point function at equal-time before moving on the unequal-time case. 
Departing from the EdS approximation, the time functions appearing in Eq.~\eqref{eq:facto}, in particular in the loop, are replaced by their exact time form, but still preserving the factorized form. 
However, in general, in the presence of species with large speed of sound or large mean free path, there is no factorizable form for the power spectrum. 
Nevertheless, one can use the following approximation, which we expect to be reasonably accurate.
We focus the discussion on the presence of massive neutrinos for definiteness.
The formula for the power spectrum, Eq.~\eqref{eq:powerspectrum}, has been shown to account for the effects from the presence of massive neutrinos at leading order ({i.e. in the so called log-enhanced contributions)}~\cite{Senatore:2017hyk}. 
To mitigate the scale dependence while keeping a practical evaluation, the power spectrum can be first evaluated at the effective redshift $z_{\rm eff}$ of the bin, and then rescaled by the scale-independent growth functions at each redshift spanned in the integration over the bin, as given by Eq.~\eqref{eq:facto} but with the replacement:
\begin{equation}\label{eq:rescale}
 D(z) \rightarrow D(z)/D(z_{\rm eff}), \qquad P_\eta(k,\mu) \rightarrow P_\eta(k, \mu, z_{\rm eff}).
 \end{equation}
Provided that the redshift bin is narrow enough and the neutrino masses are not too big, the difference of a power spectrum rescaled in such a way at redshift $z$ with a power spectrum evaluated directly at this redshift is small. 

\begin{figure}
\centering

\includegraphics[width=0.99\textwidth]{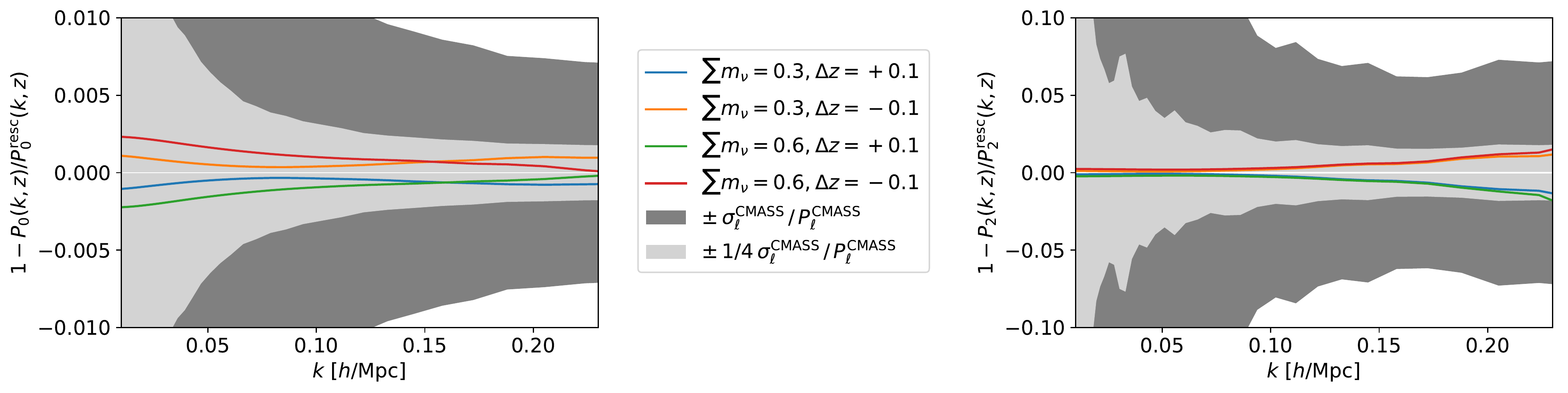}
\caption{ \small
Relative differences of the galaxy power spectrum multipoles $P_\ell(k, z)$, $\ell=0,2$, with $P_\ell^{\rm resc}(k, z)$, at redshift $z = z_{\rm eff} + \Delta z$, as a function of $k$, for various neutrino masses.
$P_\ell(k, z)$ is evaluated directly at the redshift $z$ (i.e. using the linear matter power spectrum $P_{11}(k, z)$ evaluated at $z$), while $P_\ell^{\rm resc}(k, z)$ is obtained by first evaluating at $z_{\rm eff}$ and then rescaling by $D(z)^2/D(z_{\rm eff})^2$ and $D(z)^4/D(z_{\rm eff})^4$ the linear plus counterterm and one-loop contributions, respectively. 
CMASS relative error bars are shown for comparison.
}
\label{fig:pk_nu}
\end{figure}

This is the case for BOSS samples and for the prior on the sum of neutrino masses used in this analysis, $0.06 < \sum m_\nu / {\rm eV} < 1.5$, as illustrated in Fig.~\ref{fig:pk_nu}, where we plot the difference between a power spectrum evaluated at $z=z_{\rm eff} \pm \Delta z$ and the corresponding rescaled one, for $\sum m_\nu = 0.3, 0.6 \, {\rm eV}$.
We choose $\Delta z = 0.1$, which is roughly more than the standard deviation of the galaxy number counts distribution over the CMASS sample. 
Compared to the error bars of the data, the difference is safely negligible. 
Note that the relative errors are opposite in sign for opposite departures $\pm \Delta z$ around $z_{\rm eff}$. Thus, the overall error is even less once the integration along the redshift bin is performed.
Given that the difference we find here is small, the loop that is computed approximately with the scale-independent growth introduces a negligible error. 

\paragraph{Unequal time} At unequal times, there are two differences with the previous calculation. 
First, the time functions evaluated at a single redshift $z$ are replaced by the ones at the two redshifts $z_1$ and $z_2$ for each diagram type:
\begin{align}
1-1: \, & D(\bar z)^2 \rightarrow D(z_1)D(z_2) \, , \\
2-2: \, & D(\bar z)^4 \rightarrow D(z_1)^2D(z_2)^2 \, , \\
1-3: \, & D(\bar z)^4 \rightarrow \frac{1}{2}(D(z_1)^3D(z_2) + D(z_1)D(z_2)^3) \, ,
\end{align}
where `$1-1$' corresponds to the linear terms, while the loop diagrams of type `$2-2$' and `$1-3$' have different unequal-time dependences. 
The dependence on the growth rate $f$ is also modified accordingly.

A second modification introduced by unequal time is in the IR-resummation.
In App.~\ref{app:resum}, we derive the IR-resummation for the unequal-time two-point function. 
Here, we only quote the results and discuss their implications. 
We focus on the real-space expression as it is less cumbersome, while the redshift-space one is left to App.~\ref{app:resum}.   
Denoting by $|_N$ and $||_N$ a quantity expanded up to order~$N$, respectively resummed and not resummed, the IR-resummed power spectrum reads: 
\begin{equation}
P(\k, z_1, z_2)|_N =  \sum_{j=0}^N \int \, d^3r \, e^{-i \k \cdot \r} F||_{N-j}(\k, \r, z_1, z_2) \xi_j(\r, z_1, z_2) .
\end{equation}
where $\xi_j$ is the j-th order piece of the Eulerian (not resummed) correlation function, and: 
\begin{equation}
F||_{N-j} =  K_0 \cdot K_0^{-1} ||_{N-j}.
\end{equation}
Note that these expressions are valid both for real space and redshift space. 
In real space, rotational invariance implies:
\begin{align}
K_0(\k, \r, z_1, z_2) & = \exp \left[ -\frac{1}{2} k^i k^j A_{ij}(\r, z_1, z_2) \right] \, , \\
A_{ij}(\r,z_1, z_2) &= A_0(r, z_1, z_2) \delta_{ij} + A_2(r, z_1, z_2) \hat r_i \hat r_j \, .
\end{align}
From Eq.~\eqref{eq:k0} in the appendix, we find:
\begin{align}
A_0(r, z_1, z_2) & = \frac{2}{3} D(z_1)D(z_2) \int \frac{dp}{2\pi^2} \, P_{\rm lin,0}(p) \left[1 - j_0(pr) - j_2(pr)\right] \label{eq:A0}  \\
& \qquad \qquad \qquad \qquad \qquad \qquad + \frac{1}{3} \left[ D(z_1) - D(z_2) \right]^2 \int \frac{dp}{2\pi^2} \, P_{\rm lin,0}(p) \, ,  \nonumber \\
A_2(r, z_1, z_2) & = 2 D(z_1)D(z_2) \int \frac{dp}{2\pi^2} \, P_{\rm lin,0}(p) j_2(pr) \, ,  \label{eq:A2}
\end{align}
where $j_\ell$ is the spherical Bessel function of order $\ell$.  

For $z_1=z_2$, the last term in $A_0$, Eq.~\eqref{eq:A0}, vanishes, and we obtain the familiar expressions for the equal-time correlator~\cite{Senatore:2014via}.
Thus, the IR-resummation can be generalized from equal time to unequal time merely by changing accordingly the expressions for $A_0$ and $A_2$.
For $z_1\neq z_2$, the last term in $A_0$ is non-zero and is responsible for the damping of the power spectrum at unequal time (see also~\cite{Chisari:2019tig} for a derivation within the Zel\textquoteright{}dovitch approximation).

In Fig.~\ref{fig:pkcf}, for the CMASS redshift selection function, we show the difference between the unequal-time power spectrum $P(\k, z_1 = z_{\rm eff} - \Delta z, z_2 = z_{\rm eff}+ \Delta z)$, and the equal-time power spectrum evaluated at mean redshift $P(\k, \bar z = z_{\rm eff})$, where $\Delta z= 0.1$ and $z_{\rm eff}$ is CMASS effective redshift.   
The difference of the corresponding growth functions entering at linear level are depicted as well in black dashed line. 
We notice that, for the typical standard deviation of the CMASS redshift selection function, $\Delta z \sim 0.1$, the difference in the power spectrum is mainly coming from the difference of the growth functions in the linear contribution. 
The time dependence of the loop and of the IR-resummation contributes a subleading effect. 
In particular, the damping term plays a negligible role within the CMASS redshift bin at the scales of interest.
Overall, the total difference is relatively small, about $1/5$ of the error bar on CMASS for the monopole and less for the quadrupole.

\subsection{Results on BOSS}

\begin{figure}[h!]
\centering
\includegraphics[width=0.98\textwidth]{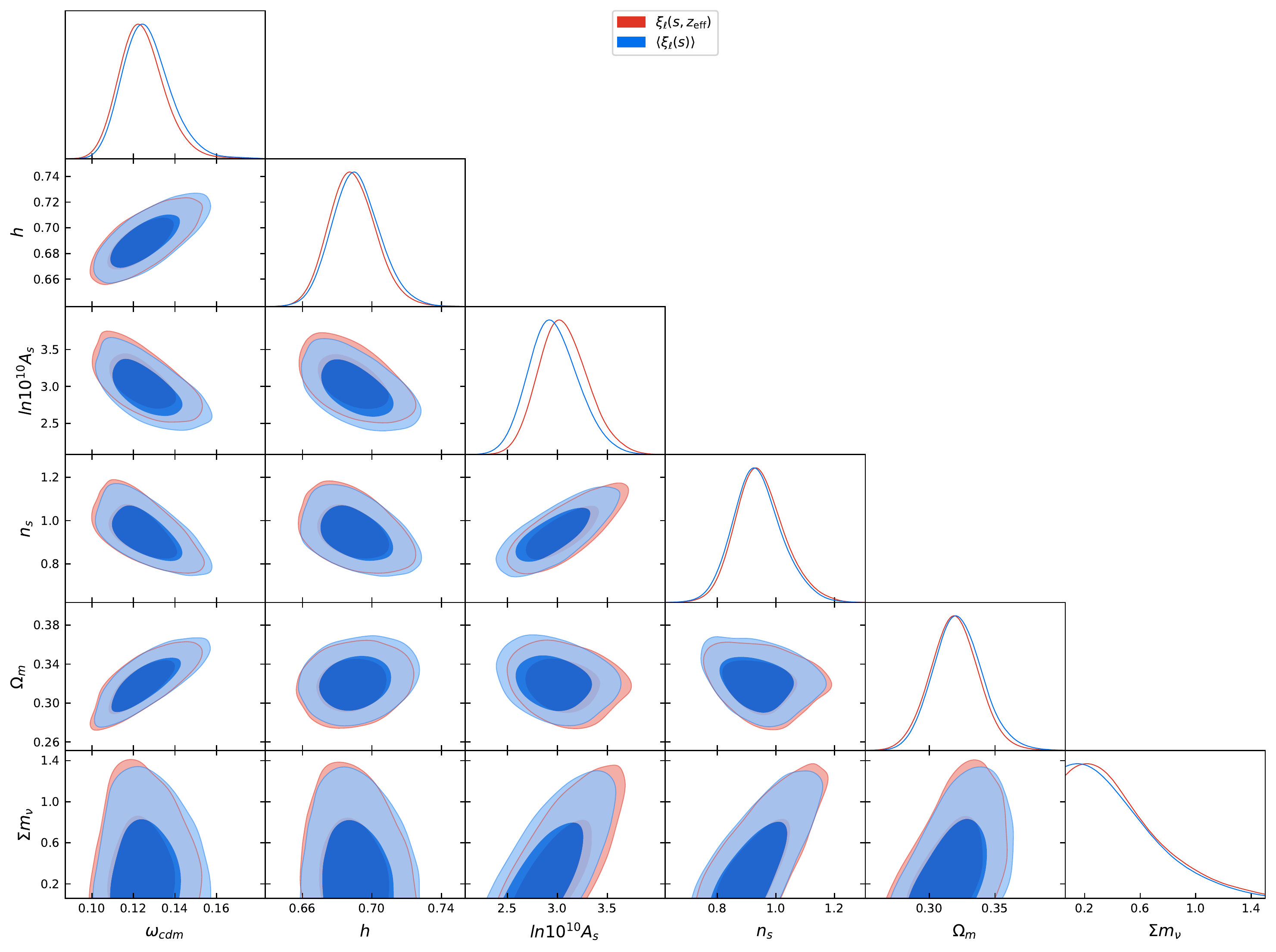}
\includegraphics[width=0.8\textwidth]{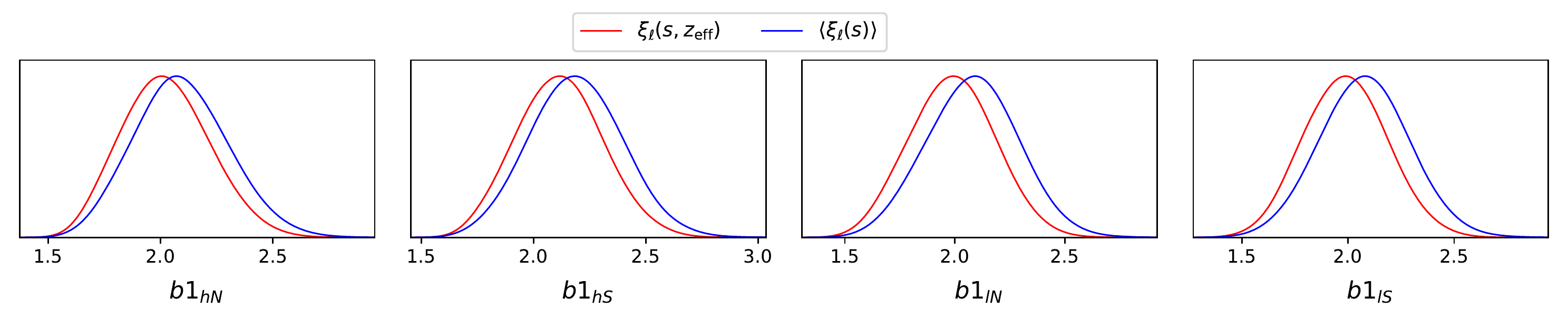}
\caption{\small Posteriors of the cosmological parameters and linear galaxy biases $b_1$ obtained by fitting the BOSS correlation function FS on $\nu\Lambda$CDM with a BBN prior, assuming either an effective redshift for each skycut or properly accounting the redshift selection functions. 
} 
\label{fig:boss}
\end{figure}

\begin{table}[]
\centering
\footnotesize
\begin{tabular}{|l|c|c|c|} 
 \hline 
$\xi_\ell(s,z_{\rm eff})$ & best-fit & mean$\pm\sigma$   \\ \hline 
$\omega_{cdm }$ &$0.1118$ & $0.1243^{+0.0092}_{-0.012} $  \\ 
$h$ &$0.678$ & $0.689^{+0.012}_{-0.014}   $  \\ 
$\ln (10^{10}A_{s })$ &$3.29$ & $3.06^{+0.22}_{-0.27}      $  \\ 
$n_{s }$ &$0.954$ & $0.947^{+0.072}_{-0.092}   $  \\ 
$\sum m_{\nu}$ [eV]  &$0.35$ & $< 1.1 (2\sigma)$  \\ 
\hline 
$\Omega_m$ & $0.300$ & $0.319\pm 0.018            $  \\
$\sigma_8$ & $0.806$ & $0.754\pm 0.055            $\\ \hline
$b_{1, \rm CMASS\, NGC }$ &$1.76$ & $2.02^{+0.18}_{-0.21}      $ \\ 
$b_{1, \rm CMASS\, SGC }$ &$1.82$ & $2.12\pm 0.19              $\\ 
$b_{1, \rm LOWZ\, NGC }$ &$1.71$ & $1.99\pm 0.19              $  \\ 
$b_{1, \rm LOWZ\, SGC }$ &$1.77$  & $1.99\pm 0.20              $ \\ 
\hline
 \end{tabular} 
\begin{tabular}{|l|c|c|c|c|} 
 \hline 
$\langle \xi_\ell(s) \rangle$ & best-fit & mean$\pm\sigma$ \\ \hline 
$\omega_{cdm }$ &$0.1217$  & $0.1266^{+0.0093}_{-0.013} $ \\ 
$h$ &$0.684$ & $0.691^{+0.013}_{-0.015}   $  \\ 
$\ln (10^{10}A_{s })$ &$3.05$ & $2.97^{+0.22}_{-0.27}      $  \\ 
$n_{s }$ &$0.930$ & $0.938^{+0.073}_{-0.090}   $ \\ 
$\sum m_{\nu}$ [eV]   &$0.35$ & $< 1.1 (2\sigma)$ \\ \hline
$\Omega_{m }$ &$0.316$ & $0.322^{+0.017}_{-0.019}   $ \\ 
$\sigma_8$ &$0.754$ & $0.728^{+0.052}_{-0.060}   $ \\ \hline
$b_{1, \rm CMASS\, NGC }$  & $1.97$& $2.09^{+0.20}_{-0.22}      $ \\ 
$b_{1, \rm CMASS\, SGC }$&$2.09$& $2.19\pm 0.21              $  \\ 
$b_{1, \rm LOWZ\, NGC }$  &$2.01$ & $2.08\pm 0.21              $ \\ 
$b_{1, \rm LOWZ\, SGC }$&$1.95$  & $2.08\pm 0.21              $ \\ 
\hline 
 \end{tabular} \\
\caption{\small Best fits and 68\%-confidence intervals of the cosmological parameters and linear galaxy biases $b_1$ obtained by fitting the FS of BOSS correlation function on $\nu\Lambda$CDM with a BBN prior, assuming either an effective redshift for each skycut or properly accounting the redshift selection function. For the total neutrino mass we quote the 95\%-confidence bound instead of the 68\%-confidence interval.}
\label{tab:boss}
\end{table}

Given that we find small but appreciable differences at the level of the observables, we verify how these translate on the determination of the cosmological parameters. 
As discussed above, we perform this check on the analysis of the correlation function in configuration space.  
The posteriors for the cosmological parameters and linear galaxy biases $b_1$ are shown in Fig.~\ref{fig:boss} and Table~\ref{tab:boss}. 
Details on the likelihood and priors are given in App.~\ref{app:eftparam}. 

In terms of the $68\%$-confidence intervals, the shifts in the posteriors of the cosmological parameters are $0.15, 0.22, 0.15, 0.39, 0.12$, for $\Omega_m$, $\omega_{cdm}$, $h$, $\ln (10^{10} A_s)$, and $n_s$, respectively. 
Thus, with the exception of a marginally small shift of $\sim 0.4 \sigma$ on $\ln (10^{10} A_s)$, the shifts are relatively small for all cosmological parameters. 
This validates the use of the effective redshift approximation for BOSS analysis as it modifies only marginally the determination of the cosmological parameters.  
We notice that the linear biases $b_1$'s are shifted relatively to about $0.34, 0.36,  0.46, 0.43$ for respectively CMASS NGC, CMASS SGC, LOWZ NGC, and LOWZ SGC. 
The difference in the observables visible in Fig.~\ref{fig:pkcf}, of $\sim 1/2 \sigma$, is thus mainly absorbed by a shift in $b_1$'s and $\ln (10^{10}A_s)$ at the level of the posteriors.

\section{Conclusion}\label{sec:conclusion}
In this work, we have checked the accuracy of two commonly used approximations to compute the galaxy power spectrum: the EdS time approximation in the one loop, and the effective redshift approximation given an observational redshift bin. 
To do so, we have re-analyzed the BOSS data following~\cite{DAmico:2019fhj,Colas:2019ret,DAmico:2020kxu} but removing these approximations: we have instead used the galaxy power spectrum with exact time dependence, and we have accounted for the redshift selection by properly integrating the power spectrum over the redshift bin. 
On the way, we have derived the one-loop galaxy power spectrum in redshift space for unequal times, including the IR-resummation. 
In summary, we find similar contours for the cosmological parameters, validating the accuracy of the time approximations made in previous BOSS FS analyses~\cite{DAmico:2019fhj,Colas:2019ret,DAmico:2020kxu,DAmico:2020ods}. 

Fitting BOSS data with a BBN prior, we find no difference in the posteriors of the cosmological parameters with or without the use of the EdS time approximation on $\Lambda$CDM with massive neutrinos in normal hierarchy with a flat prior of $0.06 < \Sigma m_\nu / {\rm eV} < 1.5$.
The same conclusion holds for simulations of volume about $50 {\rm Gpc}^3$. 
This first result shows that the EdS time approximation is good enough for upcoming surveys like DESI or Euclid. 
However, accounting for the exact time dependence of the one-loop galaxy power spectrum in redshift space does not present any particular issue compared to the EdS evaluation, except for a negligible slowdown of computation time~\cite{Donath:2020abv}. 
As such, the exact time dependence evaluation can be made standard. 

Second, the difference in the cosmological parameters obtained fitting BOSS data by properly taking into account the galaxy distribution over the redshift bin rather than using an effective redshift for each BOSS sample, is found to be relatively small, with at most $\sim 0.4 \sigma$ relative shift in~$\ln (10^{10} A_s)$. 
As the cosmological results are marginally modified, the use of the effective redshift approximation for BOSS analysis is validated. 
It is not clear if such conclusion will hold for future surveys. 
Indeed, if the relative shifts in $\ln(10^{10}A_s)$, and in the linear galaxy biases $b_1$'s, of $\sim 0.4$, are marginally significant for BOSS, larger shifts can be expected given larger data volume.  
This warns us to carefully select redshift samples. 

An intriguing avenue that the formulas in this paper allow is the analysis of the whole BOSS data, or other LSS surveys data, in one single wide redshift bin, rather than cut into two or several narrow redshift bins. 
In principle, such analysis will bring extra statistical power, as more correlations are effectively considered. 
However, this strongly depends on the assumption that the EFT parameters are evolving sufficiently mildly along the redshift bin.
If taken too wide, this assumption will necessarily break down~\footnote{Based on the discussions of Section~\ref{sec:redshiftselection}, if one consider the BOSS data in one single redshift bin, the variation of $b_1$ around effective redshift $z \sim 0.43$ to $[z_{\rm min}, z_{\rm max}] = [0.2, 0.7]$ will be then about $\pm 20\%$. 
This represents a large variation compared to the size of the error bars of $b_1$ (see Table~\ref{tab:boss}), especially since the error bars will be even smaller in the case of a single redshift bin.  
Therefore, at this level, it might become important to take into account the time evolution of the galaxy biases rather than approximating them to be constant along the bin. }.
Another possible route is to keep several separate narrow redshift bins, allowing for different values of the EFT parameters at each redshift bin, but to also analyze the cross-redshift bin signal. 
As such, we would stay safe and agnostic on the time evolution of the EFT parameters, allowing them to take different values at different redshifts, but still gain in statistics. 
We leave such explorations to future work. 

Lastly, we caution that if in $w$CDM and $\Lambda$CDM no differences are visible when accounting exactly for the time dependence, this may not be the case for other cosmologies (see e.g.~\cite{Cusin:2017wjg,Li:2018ixg}). 
As a concrete example, in an exploration of clustering quintessence with BOSS FS using the EFTofLSS, the difference from the EdS approximation is found to have non-negligible impact on the determination of the cosmological parameters~\cite{DAmico:2020tty}.  
In general, it appears necessary to assess all potential sources of systematic errors, especially given the precision in the measurements of the cosmological parameters brought by the EFTofLSS. 
On the data side, it might be worthwhile to look for potential unknown systematics such as selection effects or undetected foregrounds that can affect the results. 
We leave such investigations to future work.

\section*{Acknowledgements}
PZ would like to thank Emanuele Castorina and Jia Liu for discussions, Yaniv Donath for collaboration during the early stages of this project, and Guido D'Amico and Leonardo Senatore for valuable help and comments on the draft. 
\noindent PZ is grateful for support from the ANSO/CAS-TWAS Scholarship. 
YFC is supported in part by the NSFC (Nos. 11722327, 11653002, 11961131007, 11421303), by the CAST-YESS (2016QNRC001), by CAS project for young scientists in basic research (YSBR-006), by the National Youth Talents Program of China, and by the Fundamental Research Funds for Central Universities. 

\noindent Part numerics are operated on the computer clusters LINDA \& JUDY in the particle cosmology group at USTC, and part on the Sherlock cluster at the Stanford University.

\appendix

\section{EFT expansion}\label{app:bias}

The EFT expansion with exact time dependence has been derived up to third-order for $\Lambda$CDM in~\cite{Donath:2020abv}. 
Here, we give the formulas for $w$CDM and in the basis of descendants (BoD)~\cite{Angulo:2015eqa,Fujita:2016dne}. 

\subsection{Green's and time functions}
The growth factor is defined as the solution of:
\be \label{delta-m34}
	\frac{d^2}{d\ln a^2}D(a)+\left(2+\frac{d\ln H}{d\ln a}\right)\frac{d}{d\ln a}D(a)-\frac{3}{2}\Omega_{m}(a) D(a)=0 \, ,
\ee
which results from the linear equations of motion of the smoothed fields. 
The Hubble parameter reads:
\be
	H(a) = H_0\sqrt{\Omega_{m,0}a^{-3}+\Omega_{D,0}a^{-3(1+w)}} \, ,
\ee
where the fractional matter and dark energy densities are given by:
\be
\Omega_m(a) = \Omega_{m,0}\frac{H_0^2}{H(a)^2}a^{-3} \quad\text{and}\quad \Omega_D(a) = \Omega_{D,0}\frac{H_0^2}{H(a)^2}a^{-3(1+w)},
\ee
and  $\Omega_{m,0}$ and $\Omega_{D,0}$ are their present day values.
There are two solutions to Eq.~\eqref{delta-m34}, which are given in terms of hypergeometric functions~\cite{Lee:2009gb}. 
The growing mode reads:
\be
D_{+}(a)=a\cdot\,{}_2F_1\left(\frac{w-1}{2w}, -\frac{1}{3w}, 
1 -\frac{5}{6w} , -a^{-3 w} \frac{\Omega_{D,0}}{\Omega_{m,0}}\right) \, ,
\ee
and the decaying mode is:
\be
D_{-}(a)=a^{-\frac{3}{2}}\cdot\,{}_2F_1\left(\frac{1}{2w}, \frac{1}{2}+\frac{1}{3w}, 
1 +\frac{5}{6w} , -a^{-3 w} \frac{\Omega_{D,0}}{\Omega_{m,0}}\right) \, .
\ee
We construct the solutions with exact time dependence by defining the following Green's functions from the equations of motion: 
\begin{align}
&a \frac{d G^{\delta}_{\sigma}(a,\ta)}{da}-f_{+}(a)G^{\theta}_{\sigma}(a,\ta)=\lambda_{\sigma}\delta_D(a-\ta), \label{Green} \\
&a \frac{d G^{\theta}_{\sigma}(a,\ta)}{da}-f_{+}(a)G^{\theta}_{\sigma}(a,\ta)+\frac{3}{2} \frac{\Omega_{m}}{f_{+}}\left(G^{\theta}_{\sigma}(a,\ta)-G^{\delta}_{\sigma}(a,\ta)\right)=(1-\lambda_{\sigma})\delta_D(a-\ta),
\end{align}
where {$\sigma \in \{1,2\}$ and} $\lambda_1 = 1$ and $\lambda_2 = 0$, and $\delta_D$ denotes the Dirac delta distribution. 
We can write them explicitly as:
\begin{align}
&G^{\delta}_1(a,\ta)=\frac{1}{\ta W(\ta)}\bigg(\frac{d D_{-}(\ta)}{d\ta}D_{+}(a)-\frac{d D_{+}(\ta)}{d\ta}D_{-}(a)\bigg) {\Theta}(a-\ta) \label{gdelta} \ ,\\
&G^{\delta}_2(a,\ta)=\frac{f_{+}(\ta)/\ta^2}{W(\ta)}\bigg(D_{+}(\ta)D_{-}(a)-D_{-}(\ta)D_{+}(a)\bigg){\Theta}(a-\ta) \ , \\
&G^{\theta}_1(a,\ta)=\frac{a/\ta}{f_{+}(a)W(\ta)}\bigg(\frac{d D_{-}(\ta)}{d\ta}\frac{d D_{+}(a)}{d a}-\frac{d D_{+}(\ta)}{d\ta}\frac{d D_{-}(a)}{d a}\bigg) {\Theta}(a-\ta) \ ,\\
&G^{\theta}_2(a,\ta)=\frac{f_{+}(\ta)a/\ta^2}{f_{+}(a)W(\ta)}\bigg(D_{+}(\ta)\frac{d D_{-}(a)}{d a}-D_{-}(\ta)\frac{d D_{+}(a)}{d a}\bigg) {\Theta}(a-\ta) \ , \label{gtheta}
\end{align}
where $W(\ta)$ is the Wronskian of $D_+$ and $D_-$ and
$\Theta (a-\tilde a)$ is the Heaviside step function. We impose the boundary conditions:
\begin{align} \label{bound}
& G^{\delta}_\sigma(a,\tilde a) = 0 \quad \quad \text{and} \quad\quad G^{\theta}_\sigma(a, \tilde a)=0 \quad \quad \text{for} \quad \quad \tilde a > a \ , \\
&G^\delta_\sigma ( \tilde a , \tilde a ) = \frac{\lambda_\sigma}{\tilde a} \quad \hspace{.06in} \text{and} \hspace{.2in} \quad G^{\theta}_{\sigma} ( \tilde a , \tilde a ) = \frac{(1 - \lambda_\sigma)}{\tilde a} . \label{bound2}
\end{align}
The time functions that we need are then given by:
\begin{align} 
&\mG^{\lambda}_{\sigma}(a)=\int^{1}_0 G^{\lambda}_{\sigma}(a,\ta)\frac{f_{+}(\ta) D_{+}^2(\ta) }{ D_+^2 (a) }d\ta \ , \\ \nonumber
&\mU^{\lambda}_{\sigma}(a)=\int^{1}_0 G^{\lambda}_{1}(a,\ta)\frac{f_{+}(\ta)D^3_{+}(\ta)}{ D^3_+ (a) }\mG^{\delta}_{\sigma}(\ta)d\ta, \quad \mV^{\lambda}_{\sigma\tilde\sigma}(a)=\int^{1}_0 G^{\lambda}_{\tilde\sigma}(a,\ta)\frac{f_{+}(\ta)D^3_{+}(\ta)}{ D^3_+ (a) }\mG^{\theta}_{\sigma}(\ta)d\ta, \label{eq:timefunc}
\end{align}
where $\lambda \in \{\delta,\theta\}$ and $\sigma,\tilde{\sigma} \in \{1,2\}$.

The additional function entering the bias derivative expansion, Eq.~\eqref{eq:density}, is given by:
\bea
Y(a) = -\frac{3}{14}+\mV^{\delta}_{11}(a)+\mV^{\delta}_{12}(a).
\eea

\subsection{Galaxy kernels}

		The biased density and velocity divergence read, in the BoD:
		\bea \label{eq:delta_h_CoI_t_bod}
		\delta_h(\vec k,t) & = & \tilde{c}_{\delta,1}(t) \; \Big( \mathbb{C}^{(1)}_{\delta, 1}(\vec k, t )+ \mathbb{C}^{(2)}_{\delta, 1}(\vec k, t) + \mathbb{C}^{(3)}_{\delta, 1}(\vec k, t) +Y(a)\mathbb{C}^{(3)}_{Y}(\vec k,a)\Big) 
		\\
		& + & \tilde{c}_{\delta, 2}(t) \; \Big( \mathbb{C}^{(2)}_{\delta, 2}(\vec k, t) + \mathbb{C}^{(3)}_{\delta, 2}(\vec k, t) \Big) \nonumber
		\\
		& + & \tilde{c}_{\delta^2, 1}(t) \; \Big( \mathbb{C}^{(2)}_{\delta^2, 1}(\vec k, t) + \mathbb{C}^{(3)}_{\delta^2, 1}(\vec k, t) \Big) \nonumber
		\\
		& + & \tilde{c}_{\delta, 3}(t) \; \mathbb{C}^{(3)}_{\delta, 3}(\vec k, t)+\tilde{c}_{\delta^2, 2}(t) \; \mathbb{C}^{(3)}_{\delta^2 ,2}(\vec k, t)
		 \nonumber
		\\
		& + &   \tilde{c}_{s^2, 2}(t) \; \mathbb{C}^{(3)}_{s^2, 2}(\vec k, t)+\tilde{c}_{\delta^3}(t) \; \mathbb{C}^{(3)}_{\delta^3}(\vec k, t) \ , \nonumber
\eea

\bea \label{eq:delta_h_CoI_t}\theta_h(\vec k,t) & = & \; \Big( \mathbb{C}^{(1)}_{\delta, 1}(\vec k, t )+ \mathbb{C}^{(2)}_{\delta, 1}(\vec k, t) + \mathbb{C}^{(3)}_{\delta, 1}(\vec k, t) +Y(a)\mathbb{C}^{(3)}_{Y}(\vec k,a)\Big) 
\\
& + & \left(\frac{7}{2} - \frac{7}{2}\mG_1^\theta\right) \; \Big( \mathbb{C}^{(2)}_{\delta, 2}(\vec k, t) + \mathbb{C}^{(3)}_{\delta, 2}(\vec k, t) \Big) \nonumber
\\
& + &\left(-\frac{5}{2} + \frac{7}{2}\mG_1^\theta\right) \; \Big( \mathbb{C}^{(2)}_{\delta^2, 1}(\vec k, t) + \mathbb{C}^{(3)}_{\delta^2, 1}(\vec k, t) \Big) \nonumber
\\
& + & \left(\frac{45}{7} - 9 \mV_{12}^\theta - \frac{45}{2} \mV_{21}^\theta\right) \; \mathbb{C}^{(3)}_{\delta, 3}(\vec k, t)+\left(-\frac{22}{7} - \frac{7}{4} \mG_1^\theta + \frac{25}{6} \mV_{12}^\theta + \frac{151}{8} \mV_{21}^\theta\right) \; \mathbb{C}^{(3)}_{\delta^2 ,2}(\vec k, t)
\nonumber
\\
& + &   \left(-\frac{3}{7} + 2 \mV_{12}^\theta + \frac{3}{2} \mV_{21}^\theta)\right) \; \mathbb{C}^{(3)}_{s^2, 2}(\vec k, t)+\left(\frac{10}{7} + \frac{7}{2}\mG_1^{\theta}- 2 \mV_{12}^\theta - \frac{69}{4} \mV_{21}^\theta)\right) \; \mathbb{C}^{(3)}_{\delta^3}(\vec k, t) \ , \nonumber
			\eea
where the expressions for the operators $\mathbb{C}$ can be found in~\cite{Fujita:2016dne}, and $\mathbb{C}^{(3)}_Y$ in~\cite{Donath:2020abv}. 
 
The galaxy kernels read:
 \begin{align}
     K_1 & = b_1, \\
     K_2(\q_1,\q_2) & = b_1 \frac{\q_1\cdot \q_2 (q_1^2 + q_2^2)}{2 q_1^2 q_2^2}+ b_2\left( F_2(\q_1,\q_2) -  \frac{\q_1\cdot \q_2 (q_1^2 + q_2^2)}{2 q_1^2 q_2^2} \right) + b_4 \, , \\
     K_3(\q,-\q,\k) & = \frac{b_1}{504 k^3 q^3}\left( -38 k^5q + 48 k^3 q^3 - 18 kq^5 + 9 (k^2-q^2)^3\log \left[\frac{k-q}{k+q}\right] \right) \nonumber \\
    &+ \frac{b_3}{756 k^3 q^5} \left( 2kq(k^2+q^2)(3k^4-14k^2q^2+3q^4)+3(k^2-q^2)^4 \log \left[\frac{k-q}{k+q}\right]  \right) \nonumber \\
    & +\frac{b_1\,Y(a)}{36 k^3 q^3} \left( 6k^5 q + 16 k^3 q^3 - 6 k q^5 + 3 (k^2 - q^2)^3 \log \left[\frac{k-q}{k+q}\right] \right) \, ,
 \end{align}
where $F_2$ is the symmetrized standard perturbation theory second order density kernel (see e.g.~\cite{Bernardeau:2001qr} for explicit expressions). 
Here the third-order kernel has been UV-subtracted and we have already performed the angular integration over $x = \hat k \cdot \hat q$. 
{The galaxy biases appearing in the one loop are defined as:}
\begin{equation}
b_1 = \tilde c_{\delta,1} \, , \qquad b_2 = \tilde c_{\delta,2} \, , \qquad b_3 = \tilde c_{\delta,3} + 15 \tilde c_{s^2,2} \, , \qquad b_4 = \tilde c_{\delta^2,1} \, .
\end{equation}
The velocity kernels are then simply biased density kernels with a specific choice of biases, see Eq.~\eqref{eq:G_n} and Eq.~\eqref{eq:replacement}.

\section{Two-point function estimators}\label{app:estimator}
In order to highlight the redshift dependence of the observables, we here give a derivation of the expectation values of the two commonly-used estimators of the two-point function: 
the \emph{Landy \& Szalay estimator} for the configuration-space correlation function, and the \emph{Yamamoto estimator} for the Fourier-space power spectrum. 

\subsection{Correlation function}

Landy \& Szalay constructed an estimator for the correlation whose variance was shown to approach Poisson~\cite{1993ApJ...412...64L}. 
It is nowadays routinely used.
We will focus the discussion on this one, although it is not important for what we will show which estimator is used: 
at the level of their expectation value, all estimators are equivalent. 

\paragraph{Estimator}
The correlation function measured using the Landy \& Szalay estimator is defined as:
\begin{equation}\label{eq:estimator}
\hat \xi = \frac{DD - 2DR + RR}{RR} \, ,
\end{equation}
where $DD$, $DR$, and $RR$ are the sums over the pair counts drawn from the data catalog, data-random cross-catalog, and random catalog, respectively. 
They are given by:
\begin{align}
DD(\s) &= \sum_{ij} \delta^{(3)}_D(\s + \r_1  - \r_2) w(\r_1) w(\r_2) n^d_{i}(\r_1) n^{d}_j(\r_2) \, , \label{eq:dd}\\
DR(\s) &= \alpha \sum_{ij}\delta^{(3)}_D(\s + \r_1 - \r_2)  w(\r_1) w(\r_2) n^d_i(\r_1) n^r_j(\r_2) \, , \\
RR(\s) &= \alpha^2 \sum_{ij}\delta^{(3)}_D(\s + \r_1  - \r_2) w(\r_1) w(\r_2) n^r_i(\r_1) n^r_j(\r_2) \, , \label{eq:rr}
\end{align}
where $i \neq j$ and $ \delta^{(3)}_D$ is the Dirac distribution. 
We denote $\s \equiv \r_2 - \r_1$ the (comoving) separation vector between two objects at position $\r_1$ and $\r_2$. 
Here $n^d_i(\r) \equiv \delta^{(3)}_D(\r - \r_i)$ and $n^r_i(\r) \equiv \alpha^{-1} \delta^{(3)}_D(\r - \r_i)$ are simply the number counts of an observed object $i$ or random $i$, respectively, within a volume $dr^3$ centered on $\r$, and $w(\r)$ is a weight chosen by the observer. 
The number of randoms is taken to be much higher by a factor $\alpha^{-1}$ than the observed number counts to make the variance of the randoms negligible. 
The factors of $\alpha$ appearing in $DR$ and $RR$ will rescale the random number density to the observed one. 

\paragraph{Expectation value} Since within a redshift slice we expect homogeneity, the expectation values of the number density can be written as:
\begin{equation}
\braket{n^d_i(\r)} = \alpha^{-1} \braket{n^r_i(\r)}  =  \Theta(\r) \bar n(r) \, ,
\end{equation}
where $\bar n(r)$ is the radial selection function, and $\Theta(\r)$ is 1 if $\r$ falls within the observed survey volume, and 0 if it is outside. 
The radial selection function is typically given by the observed mean number density per redshift slice.
The correlations then are, by definition:
\begin{align}
\braket{n^d_i(\r_1) n^d_j(\r_2)} & = \Theta(\r_1) \Theta(\r_2) \bar n(r_1) \bar n(r_2)  \left( 1 + \xi(\r_1, \r_2) \right)\, , \label{eq:corr1} \\
\alpha^{-2} \braket{n^r_i(\r_1) n^r_j(\r_2)} & =  \alpha^{-1} \braket{n^d_i(\r_1) n^r_j(\r_2)} =\Theta(\r_1) \Theta(\r_2)  \bar n(r_1) \bar n(r_2) \, , \label{eq:corr2}
\end{align}
for $i \neq j$, in the limit where the randoms catalog has no noise. 

To evaluate the expectation value of the pair counts, Eq.~\eqref{eq:dd}, we can take the continuous limit such as:
\begin{align}
\braket{ DD(\s) } & =  \sum_{ij} \delta^{(3)}_D(\s + \r_1  - \r_2) w(\r_1) w(\r_2) \braket{ n^d_i(\r_1) n^d_j(\r_2) } \\
& =  \sum_{ij} \delta^{(3)}_D(\s + \r_1  - \r_2) \Theta(\r_1) \Theta(\r_2) \bar n(r_1) \bar n(r_2) \left( 1 + \xi(\r_1, \r_2) \right) \\ 
& \rightarrow \int dr_1^3 dr_2^3 \, \delta^{(3)}_D(\s + \r_1  - \r_2) \Theta(\r_1) \Theta(\r_2) \bar n(r_1) \bar n(r_2) \left( 1 + \xi(\r_1, \r_2) \right) \\
& = \int dr_1^3 \, \Theta(\r_1) \Theta(\s + \r_1) \bar n(r_1) \bar n(r_2(\s, \r_1)) \left( 1 + \xi(\s, \r_1) \right) \\
& \equiv G(\s)  \int dr_1^3 \, \bar n(r_1) \bar n(r_2(\s, \r_1)) \left( 1 + \xi(\s, \r_1) \right)   \, ,
\end{align}
where $G(\s)$ is the fraction of pairs with separation $\s$ within the survey volume~\footnote{One can think of the equality to the last line as she is on her way to Monte Carlo.}. 
Here we have redefined at the second line $\bar n(r) \equiv w(r) \bar n(r)$ for conciseness. 
In general, the weights might depend on $\r$ instead of $r$, however, as far as we are concerned with the FKP weight (see Sec.~\ref{sec:ps}), it depends only on the redshift. 

Similarly, we obtain:
\begin{equation}
\braket{ DR(\s) } = \braket{ RR(\s) }  = G(\s)  \int dr_1^3 \, \bar n(r_1) \bar n(r_2(\s, \r_1)) \, . 
\end{equation}
The expectation value of the estimator~\eqref{eq:estimator} follows:
\begin{equation}\label{eq:mean_xi}
\braket{ \hat \xi(\s) } = \frac{1}{N_\xi(\s)} \int dr_1^3 \, \bar n(r_1) \bar n(r_2(\s, \r_1)) \xi(\s, \r_1)  \, , \quad N_\xi(\s) =  \int dr_1^3 \, \bar n(r_1) \bar n(r_2(\s, \r_1)) \, ,
\end{equation}
where the weight from the survey geometry $G(\s)$ nicely cancels out. 

When using a pair-count estimator, the line of sight is often chosen as the mean direction of the pair, $\r \equiv \frac{1}{2}(\r_1 + \r_2)$. 
With such choice, the two objects in the pair are symmetric around the line of sight and thus the `wide-angle' corrections from the two true line of sights starts at $\sim \mathcal{O}\left((s/r)^2\right)$  
(see e.g.~\cite{Szalay:1997cc,Reimberg:2015jma,Castorina:2017inr}). 
Changing the integration variable $\r_1 \rightarrow \r$, we finally obtain:
\begin{equation}\label{eq:mean_xi_los}
\braket{ \hat \xi(s) } = \frac{2\pi}{N_\xi(s)} \int dr \, r^2 \int_{-1}^{+1} d\mu \, \bar n(r_1) \bar n(r_2) \xi(s, \mu, r)  \, , \quad N_\xi(s) =  2\pi \int dr \, r^2 \int_{-1}^{+1} d\mu \, \bar n(r_1) \bar n(r_2) \, ,
\end{equation}
where $r_{1,2} \equiv \sqrt{r^2 + (s/2)^2 \mp s r \mu}$, and $\mu \equiv \hat s \cdot \hat r$. 

\paragraph{Plane-parallel approximation}
For computational reason, the two objects are often taken to be at the same radial distance: $r_1 = r = r_2$, i.e. same redshift $z_1 = z = z_2$. 
In this so-called plane-parallel limit, 
\begin{equation}\label{eq:plane-parallel}
\braket{ \hat \xi(s) } \simeq \frac{4\pi}{N_\xi} \int dr \, r^2 \, \bar n(z(r))^2 \xi(s, z(r))  \, , \quad N_\xi =  4\pi \int dr \, r^2 \, \bar n(z(r))^2 \, .
\end{equation}
How good is this approximation is discussed below in~\ref{app:planeparallel}.  

\paragraph{Redshift space}
The generalization to redshift space reads:
\begin{equation}
\hat \xi_\ell(\s) =  (2\ell + 1) \frac{ \sum_{ij}  \delta^{(3)}_D(\s + \r_1  - \r_2) \big[ n^d_i(\r_1) n^d_j(\r_2)  - 2 n^d_i(\r_1) n^r_j(\r_2) + n^r_i(\r_1) n^r_j(\r_2) \big] \mathcal{L}_\ell(\mu_{ij})}{\sum_{ij} \delta^{(3)}_D(\s + \r_1  - \r_2) n^r_i(\r_1) n^r_j(\r_2)} \, ,
\end{equation}
where $\mu_{ij}$ is the cosine of the pair $ij$ in the direction of the line of sight and $\mathcal{L}_\ell$ is the Legendre polynomial of order $\ell$. 
The derivation of the expectation value is similar as in real space and leads to:
\begin{align}\label{eq:red_estimator}
\braket{ \hat \xi_\ell(s) } & = \frac{2\pi (2\ell + 1)}{N_\xi(s)} \int dr \, r^2 \int_{-1}^{+1} d\mu \, \bar n(r_1) \bar n(r_2) \xi(s, \mu, r) \mathcal{L}_\ell(\mu) \\
& = \frac{2\pi (2\ell + 1)}{N_\xi(s)} \int dr \, r^2 \int_{-1}^{+1} d\mu \, \bar n(r_1) \bar n(r_2) \sum_{\ell'} \xi_{\ell'}(s,\mu, r) \mathcal{L}_{\ell'}(\mu) \mathcal{L}_\ell(\mu)  \, ,
\end{align}
where $r_{1,2} \equiv \sqrt{r^2 + (s/2)^2 \mp s r \mu}$, and $\mu \equiv \hat s \cdot \hat r$. 

In the plane-parallel limit, $\xi_{\ell'}(s,\mu, r) \simeq \xi_{\ell'}(s, r)$ since $s_1 \simeq s_2$. 
Using the orthogonality identity of the Legendre polynomials,
\begin{equation}\label{eq:legendre_orthogonality}
\frac{2\ell+1}{2} \int d\mu \, \mathcal{L}_{\ell'}(\mu) \mathcal{L}_\ell(\mu)  = \delta_{\ell\ell'} \, ,
\end{equation}
the plane-parallel limit then reads:
\begin{equation}
\braket{ \hat \xi_\ell(s) } \simeq \frac{4\pi}{N_\xi} \int dr \, r^2 \, \bar n(z(r))^2 \xi_\ell(s, z(r)) \, .
\end{equation}

\subsection{Power spectrum}\label{sec:ps}

Feldman, Kaiser and Peacock (FKP) constructed an estimator for the power spectrum for which the variance is controlled by an optimal weighting scheme~\cite{Feldman:1993ky}. 
Here we follow closely their derivation, and then present the generalization to all multipoles in redshift space proposed by Yamamoto \emph{et al.}~\cite{Yamamoto:2005dz}. 

\paragraph{FKP estimator} We first define the weighted galaxy fluctuation field,
\begin{equation}\label{eq:F}
F(\r) = \frac{w(\r)}{N_P^{1/2}} \left[ n^d(\r) - \alpha n^r(\r) \right] \, . 
\end{equation}
where $N_P \equiv \int d^3 r \, w(\r)^2 \bar n(\r)^2$ is a convenient normalisation, and $w(\r)$ the weight. 
Here $n^d(\r) \equiv \sum_i n^d_i(\r)$ and $n^r(\r) \equiv \sum_i n^r_i(\r)$. 
From Eqs.~\eqref{eq:corr1}~and~\eqref{eq:corr2}, one can figure that their correlations are (see also App.~A of~\cite{Feldman:1993ky}):
\begin{align}
\braket{n^d(\r_1)n^d(\r_2)} & = \Theta(\r_1) \Theta(\r_2) \bar n(r_1) \bar n(r_2) \left( 1 + \xi(\r_1, \r_2) \right) +  \Theta(\r_1) \bar n(r_1) \delta(\r_1 - \r_2) \, ,\\
\braket{n^d(\r_1)n^r(\r_2)} & = \alpha^{-1} \Theta(\r_1) \Theta(\r_2) \bar n(r_1) \bar n(r_2) \, ,\\
\braket{n^r(\r_1)n^r(\r_2)} & = \alpha^{-2} \Theta(\r_1) \Theta(\r_2) \bar n(r_1) \bar n(r_2) +  \alpha^{-1} \Theta(\r_1) \bar n(r_1) \delta(\r_1 - \r_2) \, .
\end{align}
We can now evaluate the expectation value of the Fourier transform of $F(\r)$ squared:
\begin{align}
\braket{|F(\k)|^2} & = \frac{1}{N_P}  \int d^3 r_1 \, d^3 r_2 \, e^{- i \k \cdot (\r_2 - \r_1)} w(\r_1) w(\r_2) \braket{(n^d(\r_1)-\alpha n^r(\r_1))(n^d(\r_2)-\alpha n^r(\r_2))} \\
	& = \frac{1}{N_P}  \int d^3 r_1 \, d^3 r_2 \, e^{- i \k \cdot (\r_2 - \r_1)} \Theta(\r_1) \Theta(\r_2) \bar n(r_1) \bar n(r_2) \xi(\r_1, \r_2) + \frac{1+\alpha}{N_P} \int d^3 r \, \Theta(\r) w(r) \bar n(r) \, ,
\end{align}
where we have redefined $\bar n(r) \equiv w(r) \bar n(r)$ at the second line. 

The FKP estimator is defined as the average over a shell in $k$-space of the raw power spectrum shot-noise subtracted:
\begin{equation}\label{eq:kfp}
 \hat P(k)  \equiv \int \frac{d\Omega_k}{4\pi} \left[ |F(\k)|^2  - P^{\rm noise} \right] \, ,
\end{equation}
where $P^{\rm noise}  \equiv  \frac{1+\alpha}{N_P} \int d^3 r \, \Theta(\r) \bar n(r)$. 

The optimal weight that minimizes the variance of the estimator is, in the limit of Gaussian long-wavelength fluctuations,
\begin{equation}\label{eq:fkp_weight}
w(r) = \frac{1}{1+\bar n(r) P(k)} \,
\end{equation}
where $P(k)$ can be chosen constant for practical purpose.  

\paragraph{Yamamoto estimator} 
The Yamamoto estimator generalizes the KFP estimator to all multipoles. 
The power spectrum multipoles measured using the Yamamoto estimator is defined as:
\begin{equation}\label{eq:yamamoto}
\hat P_\ell(k) = \int \frac{d\Omega_k}{4\pi} \left[ \frac{2\ell+1}{N_P} \int d^3r_1 d^3r_2 \, e^{-i \k \cdot (\r_2 - \r_1)} F(\r_1) F(\r_2) \mathcal{L}_\ell(\hat k \cdot \hat r_1) - P_\ell^{\rm noise}(\k) \right] \, ,
\end{equation}
where $P^{\rm noise}_\ell(\k)  \equiv (2\ell+1)(1+\alpha) N_P^{-1} \int d^3 r \, \Theta(\r)  w(r) \bar n(r) \mathcal{L}_\ell(\hat k \cdot \hat r)$. 
Here the line of sight is chosen to be in the direction of one of the object in the pair, $\hat r_1$. 
This choice is motivated for computational reason: the two integrals in $\r_1$ and $\r_2$ can be performed separately, for example, using fast Fourier transforms~\cite{Bianchi:2015oia,Scoccimarro:2015bla}. 
The expectation value reads:
\begin{equation}
\braket{ \hat P_\ell(k) } =  \frac{2\ell+1}{2 N_P} \int \frac{d\Omega_k}{4\pi} d^3r_1 d^3r_2 \, e^{-i \k \cdot (\r_2 - \r_1)} \Theta(\r_1) \Theta(\r_2) \bar n(r_1) \bar n(r_2) \xi(\r_1, \r_2) \mathcal{L}_\ell(\hat k \cdot \hat r_1) \, .
\end{equation}
Let us denote $\s \equiv \r_2 - \r_1$ the (comoving) separation vector between the two objects, and $\mu \equiv \hat s \cdot \hat r_1$ its cosine with the line of sight. 
Changing the integration variable $\r_2 \rightarrow \s$, and using the following identity:
\begin{equation}
\int \frac{d\Omega_k}{4 \pi} e^{-i \k \cdot \s}  \mathcal{L}_\ell (\hat k \cdot \hat r_1) = (-i)^\ell j_\ell(ks) \mathcal{L}_\ell (\hat s \cdot \hat r_1) \, ,
\end{equation}
we obtain 
\begin{equation}\label{eq:mean_yamamoto}
\braket{ \hat P_\ell(k)} = \frac{2\pi (2\ell+1)}{N_P} (-i)^\ell \int ds \, s^2 j_\ell(ks)  \int_{-1}^{+1} d\mu \, \int dr_1 \, r_1^2 \, \bar n(r_1) \bar n(r_2)  \xi(s, \mu, r_1) \, \mathcal{L}_\ell(\mu) Q(s, \mu, r_1) \, ,
\end{equation}
where $r_2 \equiv \sqrt{s^2 + r_1^2 + 2 s r_1 \mu}$, and $Q(s, \mu, r_1)$ is the window function defined as:
\begin{equation}\label{eq:3dwin}
Q(s, \mu, r_1) \equiv \frac{1}{8\pi^2} \int d\Omega_{r_1} \int_0^{2\pi} d\phi_s \, \Theta(\r_1) \Theta(\r_2) \, .
\end{equation}
Neglecting the survey geometry, the expectation value of the FKP estimator is then simply the spherical-Bessel transform of the expectation value of the Landy \& Szalay estimator, but with a different line-of-sight definition (and slight different normalisation). 

Using Eq.~\eqref{eq:legendre_orthogonality}, the plane-parallel limit reads:
\begin{equation}\label{eq:ps_plane-parallel}
\braket{ \hat P_\ell(k)} = \frac{4\pi}{N_P} (-i)^\ell \int ds \, s^2 j_\ell(ks) \, \int dr_1 \, r_1^2 \, \bar n(r_1)^2  \xi_\ell(s, r_1) \tilde Q(s, r_1) \, ,
\end{equation}
where $\tilde Q(s, r_1) \equiv \frac{1}{8\pi^2} \int d\Omega_{r_1} \int d\Omega_{s} \, \Theta(\r_1) \Theta(\r_2)$. 

\subsection{Beyond the plane-parallel limit}\label{app:planeparallel}

Here we show that the plane-parallel approximation~\eqref{eq:plane-parallel} or~\eqref{eq:ps_plane-parallel} is valid up to $\sim \mathcal{O}( (s/r)^2 )$ corrections in real space, or for even multipoles in redshift space, where $s$ is the separation and $r$ is the mean radial distance. 
In the main text, however, we do not use the expansions written here but rather implement the more general formula derived above.

There are two common choices for the line of sight $\r$: either to be in the direction of one of the pair (end-point LOS), e.g. $\r \equiv \r_1$, either in the direction of the pair (mean LOS), $\r \equiv \frac{1}{2}(\r_1 + \r_2)$. 
Let us expand $r_1$ and $r_2$ respectively around $r$ in powers of $(s/r)$:
\begin{align}
r_1 = r, \quad r_2 = \sqrt{r^2 + s^2 + 2rs\mu} = r \left[ 1 + \left(\frac{s}{r} \right)^2 + 2 \frac{s}{r} \mu + 2\left( \frac{s}{r} \right)^2 \mu^2 \dots \right] \quad & \textrm{(end-point LOS)} \, ,\\
r_{1,2} = \sqrt{r^2 + (s/2)^2 \mp rs\mu}  = r \left[ 1 + \left(\frac{s}{2r} \right)^2 \mp \frac{s}{r} \mu + \frac{1}{2} \left(\frac{s}{r} \right)^2 \mu^2 + \dots \right] \quad & \textrm{(mean LOS)} \, .
\end{align}
This implies for a generic function $f(r_1)$:
\begin{align}
f(r_1)f(r_2) & = f(r)^2 \Bigg\lbrace 1 + \frac{d\log f(r)}{d\log r}  \left[ \left(\frac{s}{r} \right)^2 + 2 \frac{s}{r} \mu + 2\left( \frac{s}{r} \right)^2 \mu^2 \right] \qquad \qquad \qquad  \quad  \textrm{(end-point LOS)} \\
& \qquad + \left( \frac{d\log f(r)}{d\log r} \right)^2 4 \left( \frac{s}{r} \right)^2 \mu^2  + \frac{d^2\log f(r)}{d\log r^2} \left(\frac{s}{r} \right)^2 \mu^2 + \dots \Bigg\rbrace \nonumber \\
& = f(r)^2 \Bigg\lbrace 1 + 2 \left( \frac{s}{r} \right) \frac{d\log f}{d\log r} \mu + \left(\frac{s}{r} \right)^2 \left[  \frac{d\log f}{d\log r} + 4  \left( \frac{d\log f}{d\log r} \right)^2 \mu^2 + f(r) \frac{d^2\log f}{d\log r^2} \mu^2 \right]  + \dots \Bigg\rbrace \, , \nonumber \\
f(r_1)f(r_2) & = f(r)^2 \Bigg\lbrace 1 + 2 \frac{d\log f(r)}{d\log r} \left[ \left(\frac{s}{2r} \right)^2 +  \frac{1}{2} \left(\frac{s}{r} \right)^2 \mu^2 \right] \qquad \qquad \qquad \qquad  \quad \textrm{(mean LOS)} \\
& \qquad - \left( \frac{d\log f(r)}{d\log r} \right)^2 \left( \frac{s}{r} \right)^2 \mu^2 + \frac{1}{2} \frac{d^2\log f(r)}{d\log r^2} \left( \frac{s}{r} \right)^2  \mu^2 + \dots \Bigg\rbrace \nonumber \\
& = f(r)^2 \Bigg\lbrace 1 + \left(\frac{s}{r} \right)^2 \left[ \frac{d\log f}{d\log r} \left(\frac{1}{2} + \mu^2 \right) - \left( \frac{d\log f}{d\log r} \right)^2 \mu^2 + \frac{1}{2} \frac{d^2 \log f}{d\log r^2} \mu^2 \right]  + \dots \Bigg\rbrace \nonumber
\end{align}

If one chooses the end-point LOS, there will be a non-zero contribution starting at $\sim \mathcal{O}( s/r )$ to the odd multipoles coming from the corrections proportional to odd powers of $\mu$. 
For even multipoles, after performing the angular integral over $\mu$ for which terms in odd powers of $\mu$ vanish, it is apparent that the first corrections to the plane-parallel approximation are $\sim \mathcal{O}( (s/r)^2 )$. 
These corrections are of the order of the wide-angle corrections that we are neglecting when we choose a single direction for the line of sight. 

Let us provide some rough estimates of the size of those corrections. 
If we are interested in the BAO, $s_{\rm BAO} \sim 110 \textrm{Mpc}/h$ and $r(z \sim 0.5) \sim 1300 \textrm{Mpc}/h$. 
Thus, the corrections are typically of order $\sim (s_{\rm BAO}/r)^2 \sim 1\%$, which are potentially non-negligible. 
Note that the wide-angle geometric corrections are controlled instead by the maximal angle of the survey,
whereas the corrections to the selection function are controlled by the maximal selected redshift range. 
These are two distinct scales, and therefore, it is important to scrutiny over both type of corrections. 

Wide-angle effects have been discussed in e.g.~\cite{Reimberg:2015jma,Castorina:2017inr,Castorina:2018nlb,Beutler:2018vpe} and shown to be small for current surveys. 
In Section~\ref{sec:beyondzeff}, we take the other route, leaving the wide-angle corrections aside, and supplement these systematic studies by assessing the impact of the redshift evolution in the observables, going beyond the effective redshift approximation and the plane-parallel limit.

\section{Unequal-time IR-resummation}\label{app:resum}

The galaxy {EFT} expansion performed in the Eulerian frame misses the effects from the bulk displacements, that are generically of order one and therefore cannot be treated perturbatively. 
These long-wavelength displacements can be resummed order by order in a parametrically controlled way, that goes under the name of IR-resummation, as originally developed in real space~\cite{Senatore:2014via}, and extended to redshift space in~\cite{Senatore:2014vja,Lewandowski:2015ziq} (see also~\cite{Blas:2016sfa,Senatore:2017pbn,Lewandowski:2018ywf,Ivanov:2018gjr} for subsequent works). 
In this appendix, we derive in real space and redshift space the IR-resummation of the two-point function for unequal-time correlation.

\subsection{Real space}
In the Lagrangian picture, the displacement field is not expanded and therefore automatically fully accounts for the long-wavelength displacements that we wish to resum to the Eulerian perturbative expansion~\cite{Porto:2013qua}. We thus start in the Lagrangian frame. 
In real space, the position of a galaxy in Lagrangian coordinate $\x$, at redshift $z$, is given by its initial position $\q$ and the displacement $\s(\q, z)$ from its initial position:
\begin{equation}
\x(\q, z) = \q + \s(\q, z).
\end{equation}
Due to mass conservation, the galaxy overdensity is given by:
\begin{equation}
1+ \delta(\x, z) = \int \, d^3q \, \delta_D(\x-\q-\s(\q, z)),
\end{equation}
where $\delta_D$ is the Dirac $\delta$-function.
Going to Fourier space yields, for $k\neq 0$:
\begin{equation}
\delta(\k, z) = \int \, d^3q \, e^{-i \k \cdot \q - i \k \cdot \s(\q, \, z)}.
\end{equation}
Thus, the real-space power spectrum reads, for $k\neq 0$:
\begin{align}
P(k, z_1, z_2) & = \int \, d^3 q \, e^{-i \k \cdot \q} \, K(\k, \q, z_1, z_2) \, , \\
K(\k, \q, z_1, z_2) & = \left\langle e^{-i \k \cdot \left[ \s(\q_1, \, z_1)-\s(\q_2, \, z_2)\right]} \right\rangle \, ,
\end{align}
where $\q = \q_2 - \q_1$ is the Lagrangian-space separation between two galaxies. 
Using the cumulant expansion theorem yields:
\begin{equation}
K(\k, \q, z_1, z_2) = \exp \left[ \sum_{n=1}^\infty \frac{(-i)^n}{n!} \left\langle \left( \k \cdot \left[ \s(\q_1, \, z_1)-\s(\q_2, \, z_2)\right] \right)^n \right\rangle \right] \, .
\end{equation}
Only the two-point function of the displacements $K_0$ in the Taylor expansion of $K=K_0 + ...$ with linear displacements $\s(p,  z) \simeq \s_{\rm lin}(p,  z)=i\frac{\p}{p^2} \delta_{\rm lin}(p, z)$ needs to be considered for resumming the IR contributions in the EFTofLSS~\cite{Senatore:2017pbn}. 
Fourier transforming each $\s(\q_i, z_i)$, $i=1, 2,$ and using the fact that: $\langle s_i(\p,z_1) s_j(\p',z_2) \rangle = (2\pi)^3\delta_D(\p+\p')\frac{p_i p_j}{p^{4}} P_{\rm lin}(p,z_1,z_2)$, 
we find:
\begin{equation}\label{eq:k0}
K_0(\k, \q, z_1, z_2) = \exp \left[ -\frac{1}{2} \int \, \frac{d^3 p}{(2\pi)^3} \frac{(\p \cdot \k)^2}{p^4} \left( P_{\rm lin}(p, z_1)+P_{\rm lin}(p, z_2) - 2 e^{i \p \cdot \q} P_{\rm lin}(p, z_1, z_2)  \right) \right] \, ,
\end{equation}
{where $P_{\rm lin}(p,z) \equiv P_{\rm lin}(p, z, z)$}.
$K_0$ contains all the information from the bulk displacements we wish to resum. 
Denoting by $|_N$ ($||_N$) a quantity resummed (not resummed) expanded up to order $N$, we can use the standard trick to resum all IR contributions up to order $N$: 
\begin{equation}
K|_N \simeq K_0 \cdot \frac{K}{K_0} \Big|\Big|_N = \sum_{j=0}^N F||_{N-j} \cdot K_j \, ,
\end{equation}
where we have defined $F||_{N-j} = K_0 \cdot K_0^{-1} ||_{N-j}$, and $K_j$ is related to the j-th order piece of the Eulerian power spectrum by:
\begin{equation}\label{eq:kj}
P_j(\k, z_1, z_2) = \int d^3q \, e^{-i \k \cdot \q} K_j(\k, \q, z_1, z_2) \, .
\end{equation}
Thus, the IR-resummed power spectrum reads:
\begin{equation}\label{eq:resumps_kj}
P(k, z_1, z_2)|_N =  \int \, d^3 q \, e^{-i \k \cdot \q} \sum_{j=0}^N F||_{N-j}(\k, \q, z_1, z_2) \cdot K_j(\k, \q, z_1, z_2) \, .
\end{equation}
In~\cite{Senatore:2017pbn}, the IR-resummation has been conveniently expressed in configuration space.
Fourier transforming Eq.~\eqref{eq:resumps_kj}, we get:
\begin{equation}
\xi(\r, z_1, z_2)|_N =\sum_{j=0}^N \int \frac{d^3 k \, d^3q}{(2\pi)^3} e^{i\k \cdot (\r-\q)} F||_{N-j}(\k, \q, z_1, z_2) \cdot K_j(\k, \q, z_1, z_2) \, .
\end{equation}
One is then free to approximate $F||_{N-j}(\k, \q)$ by expanding around $\q \simeq \r = \x_1 - \x_2$:
since $\s(\q_i) = \s(\x_i - \s(\q_i)) \simeq \s(\x_i) - \s(\x_i) \cdot \nabla \s(\x_i)$, $i=1,2,$ one can see from Eq.~\eqref{eq:k0} that higher-order terms will involve gradients of the displacement field which are parametrically suppressed with respect to the terms that are resummed, and thus can be neglected.
This allows us to pull out from the $d^3 q$ integral $F||_{N-j}(\k, \q) \rightarrow F||_{N-j}(\k, \r)$, and using Eq.~\eqref{eq:kj}, the IR-resummed correlation function is then given by:
\begin{equation}
\xi(\r, z_1, z_2)|_N =\sum_{j=0}^N \int \frac{d^3 k}{(2\pi)^3} e^{i\k \cdot \r}  F||_{N-j}(\k, \r, z_1, z_2) P_j(\k, z_1, z_2)  \, .
\end{equation}
Using the same approximations, we obtain a similar expression for the IR-resummed power spectrum:
\begin{align}
P(k, z_1, z_2)|_N  & = \sum_{j=0}^N \int \, d^3p  \, \delta_D(\p-\k)  \int \, d^3q \, e^{-i \p \cdot \q}  F||_{N-j}(\k, \q, z_1, z_2) \cdot K_j(\p, \q, z_1, z_2) \, \nonumber \\
& {=} \sum_{j=0}^N \int \, \frac{d^3p \, d^3r}{(2\pi)^3} \, e^{i \r \cdot (\p-\k)}  F||_{N-j}(\k, \r, z_1, z_2)  \int \, d^3q \, e^{-i \p \cdot \q}  K_j(\p, \q, z_1, z_2) \nonumber \\
& = \sum_{j=0}^N \int \, \frac{d^3p \, d^3r}{(2\pi)^3} \, e^{i \r \cdot (\p-\k)}  F||_{N-j}(\k, \r, z_1, z_2) P_j({\p}, z_1, z_2) \nonumber \\
& =  \sum_{j=0}^N \int \, d^3r \, e^{-i \k \cdot \r} F||_{N-j}(\k, \r, z_1, z_2) \xi_j(\r, z_1, z_2) , \label{eq:resumps}
\end{align}
where we have replaced $F||_{N-j}(\k, \q) \rightarrow F||_{N-j}(\k, \r)$ and the delta function by its integral representation in the second line, {which is exact up to order $|_N$}.

\subsection{Redshift space}
The coordinate in redshift space $\x_r$ is related to the real space one $\x$ by:
\begin{equation}
\x_r(\q, z)= \x(\q, z) + \frac{\dot{\x}(\q, z) \cdot \hat {\eta}}{H} \hat{\eta} = \q + \s(\q, z) + \frac{\dot{\s}(\q, z) \cdot \hat{\eta}}{H} \hat{\eta} \, , 
\end{equation}
where the overdot denotes a derivative with respect to time and $\hat{\eta}$ is the unit vector in the direction of the line of sight. 
The derivation is then very similar as in real space but one needs to keep track of the projection of the 3D vectors on the directions parallel and perpendicular to the line of sight.
After some steps, one obtains the same expression for the IR-resummed power spectrum, Eq.~\eqref{eq:resumps}, but with $k \rightarrow \k$ and:
\begin{align}
& K_{0}(\k, \q, z_1, z_2) = \exp \left\lbrace -\frac{1}{2} \int \, \frac{d^3 p}{(2\pi)^3} \frac{(\p \cdot \k)^2}{p^4} \left[ P_{\rm lin}(p, z_1)+P_{\rm lin}(p, z_2) - 2 e^{i \p \cdot \q} P_{\rm lin}(p, z_1, z_2)  \right] \right. \nonumber \\
& -\frac{\k \cdot \hat{\eta}}{2} \int \, \frac{d^3 p}{(2\pi)^3} \frac{(\p \cdot \k)(\p \cdot \hat{\eta})}{p^4} \left[ f(z_1) P_{\rm lin}(p, z_1) + f(z_2)P_{\rm lin}(p, z_2) - (f(z_1)+f(z_2)) e^{i \p \cdot \q} P_{\rm lin}(p, z_1, z_2)  \right] \nonumber \\
& \left. -\frac{(\k \cdot \hat{\eta})^2}{2} \int \, \frac{d^3 p}{(2\pi)^3} \frac{(\p \cdot \hat{\eta})^2}{p^4} \left[ f(z_1)^2P_{\rm lin}(p, z_1) + f(z_2)^2P_{\rm lin}(p, z_2) - 2 f(z_1)f(z_2)e^{i \p \cdot \q} P_{\rm lin}(p, z_1, z_2)  \right]  \right\rbrace \, , \label{eq:k0red}
\end{align}
where we used the fact that: $\dot{\s}_{\rm lin} = H f \s_{\rm lin}$.

{We now perform the same manipulations as in~\cite{Lewandowski:2015ziq}}. Expanding Eq.~\eqref{eq:resumps} in multipoles and dropping the time dependence to avoid clutter, the IR-resummed power spectrum multipoles read: 
\begin{align}
P_\ell(k)|_{N} & = \sum_{j=0}^N \sum_{\ell'}  4\pi (-i)^{\ell'} \int dr \, r^ 2 \, Q_{||N-j}^{\ell \ell'}(k,r) \, \xi_{\ell'}^{j} (r) \, ,\\
\xi_{\ell'}^{j} (r) & = i^{\ell'}  \int  \frac{dp\, p^2}{2\pi^2} P_{\ell'}^{j}(p)  \, j_{\ell'}(pr) \, , \\
Q_{||N-j}^{\ell \ell'}(k,r) & = \frac{2\ell+1}{2} \int_{-1}^{1}d\mu_k \,\frac{i^{\ell'}}{4 \pi} \int d^2 \hat{r} \, e^{-i\k \cdot \r} \, F_{||N-j}(\k,\r) \mathcal{L}_\ell(\mu_k) \mathcal{L}_{\ell'}(\mu_r) \,, \label{eq:resumQ}
\end{align}
where $P_{\ell}^{j}(k)$ and $\xi_{\ell}^{j} (k)$ are the $j$-loop order pieces of the Eulerian ({\it i.e.} non-resummed) power spectrum and correlation function multipoles, respectively, and $\mu_k = (\k \cdot \hat{\eta})/k$ or $\mu_r = (\r \cdot \hat{\eta})/r$ are the cosines of the angle between $\k$  or $\r$ and the line of sight $\hat{\eta}$, respectively, and $\mathcal{L}_\ell$ is the Legendre polynomial of order $\ell$. {Here, differently that in~\cite{Lewandowski:2015ziq}, we have decided to express the IR-resummed power spectrum in terms of an integral of a kernel ($Q_{||N-j}^{\ell \ell'}(k,r)$) times that correlation function $\xi_{\ell'}^{j}(r)$ rather than the power spectrum.}

As we did for the real space in Section~\ref{sec:redshiftselection}, we can express Eq.~\eqref{eq:k0red} such that the time and $\hat r$ dependence are made explicit. First define
\begin{align}
K_0(\k, \r, z_1, z_2) & = \exp \Big\lbrace -\frac{1}{2} \big[ k^i k^j A_{ij}(\r, z_1, z_2)   \\ \nn
& \qquad \qquad \qquad  + (\k \cdot \hat\eta) k^i \hat{\eta}^j B_{ij}(\r, z_1, z_2) + (\k \cdot \hat\eta)^2 \hat{\eta}^i \hat{\eta}^j C_{ij}(\r, z_1, z_2) \big] \Big\rbrace \, ,
\end{align}
{and then, if  $X= \lbrace A, B, C \rbrace$, we have}
\begin{align}
X_{ij}(\r,z_1, z_2) &= X_0(r, z_1, z_2) \delta_{ij} + X_2(r, z_1, z_2) \hat r_i \hat r_j \, ,
\end{align}
such that:
\begin{align}
X_0(r, z_1, z_2) & = \frac{u(z_1, z_2)}{3} \int \frac{dp}{2\pi^2} \, P_{\rm lin,0}(p) \left[1 - j_0(pr) - j_2(pr)\right] + \frac{v(z_1, z_2)}{3}  \int \frac{dp}{2\pi^2} \, P_{\rm lin,0}(p) \, ,  \\
X_2(r, z_1, z_2) & = u(z_1, z_2) \int \frac{dp}{2\pi^2} \, P_{\rm lin,0}(p) j_2(pr) \, ,  \label{eq:X2}
\end{align}
where:
\begin{align}
  u(z_1, z_2) & = D(z_1) D(z_2) \times
    \begin{cases}
      2 & \text{if } X=A\\
      f(z_1)+f(z_2) & \text{if } X=B \, ,\\
      2f(z_1)f(z_2) & \text{if } X=C
    \end{cases}       \\
  v(z_1, z_2) & = 
    \begin{cases}
      \left[ D(z_1) - D(z_2) \right]^2 & \text{if } X=A\\
       f(z_1) D(z_1)^2 { + } f(z_2) D(z_2)^2 - \left[f(z_1)+f(z_2)\right] D(z_1) D(z_2) & \text{if } X=B \, .\\
      \left[ f(z_1)D(z_1) - f(z_2)D(z_2) \right]^2 & \text{if } X=C
    \end{cases}       
\end{align}
$K_0$ can be conveniently re-written as:
\begin{align}
K_{0} = \exp \left\lbrace -\frac{k^2}{2} \left[  A_0 + A_2 (\hat k \cdot \hat q)^2 + B_0 \mu_k^2 + B_2 (\hat k \cdot \hat q) \mu_k \mu_q + C_0 \mu_k^2 + C_2 \mu_k^2 \mu_q^2 \right] \right\rbrace  ,
\end{align}
which, in the limit where $z_1=z_2$, agrees with the expressions found in~\cite{Lewandowski:2015ziq}.

\section{EFT parameters and Likelihoods}\label{app:eftparam}

The likelihood and the priors used in this analysis are the same as the ones used in~\cite{DAmico:2019fhj,Colas:2019ret,DAmico:2020kxu}, and are described extensively in~\cite{DAmico:2019fhj}.
In this appendix, after a brief summary, we present the posteriors of the EFT parameters obtained fitting the power spectrum multipoles of the lettered challenge simulations, with and without the EdS approximation. 

\paragraph{Likelihoods} Some EFT parameters appear only linearly in the model, and thus at most quadratically in the log-likelihood. 
This allows us to analytically marginalize over them at the level of the likelihood. 
We call the resulting likelihood the partially-marginalized likelihood, while we call the likelihood where all parameters are left varying the non-marginalized likelihood. 
We give a quick derivation of the partially-marginalized likelihood, that we use for all the analyses presented in this work. 

The theory model can be written as a sum of terms multiplied by EFT parameters appearing linearly plus all the other terms:
\begin{equation}
    P_{\alpha} = \sum_i b_{G,i} P^i_{G,\alpha} + P_{NG,\alpha} \, ,
\end{equation}
where the index $\alpha$ runs over $k$-bins and multipoles, $b_{G,i}$ are the EFT parameters over which the marginalization is analytical, and both $P^i_{G,\al}$ and $P_{NG,\al}$ depend on the cosmological parameters and the nonlinear EFT parameters which cannot be analytically integrated out.
The non-marginalized likelihood reads:
\begin{equation}
     -2 \ln \mathcal{L} = (P_{\al} - D_{\al} ) 
     C^{-1}_{\al \bt} ( P_{\bt} - D_{\bt} )
     + b_{G, i} \sigma^{-1}_{ij} b_{G, j} - 2 \ln \Pi \, ,
\end{equation}
where $D_\al$ is the data vector, $C_{\al \bt}$ is the data covariance, and we introduced a Gaussian prior on the $b_{G, i}$ with covariance $\sigma_{ij}$, plus a generic prior $\Pi$ on the cosmological and nonlinear EFT parameters.
Collecting different powers of $b_{G,i}$, the likelihood can be written as:
\begin{equation}
    \label{eq:nonmargP}
    -2 \ln \mathcal{L} = b_{G,i} F_{2, ij} b_{G,j} - 2 b_{G,i} F_{1, i} + F_0 \, ,
\end{equation}
where:
\begin{align}
    & F_{2, ij} = P^i_{G, \al} C^{-1}_{\al \bt} P^j_{G, \bt} + \sigma^{-1}_{ij} \, , \\
    & F_{1, i} = - P^i_{G, \al} C^{-1}_{\al \bt} (P_{NG, \bt} - D_{\bt}) \, , \\
    & F_0 =  (P_{NG, \al} - D_{\al}) C^{-1}_{\al \bt} (P_{NG, \bt} - D_{\bt}) - 2 \ln \Pi \, .
\end{align}
Performing a Gaussian integral on the $b_{G, i}$, the partially-marginalized likelihood follows:
\begin{equation}
    \label{eq:margP}
    -2 \ln \mathcal{L}_{\rm marg} = - F_{1,i} F^{-1}_{2,ij} F_{1,j} + F_0 + \ln \det F_2 \, .
\end{equation}
Notice that, on the best fit, the nonlinear EFT parameters can be read off by setting the gradients of Eq.~\eqref{eq:nonmargP} to zero, yielding:
\begin{equation}\label{eq:bg}
    b_{G,i} = F^{-1}_{2,ij} F_{1,j} \, .
\end{equation}

\paragraph{Priors} Following~\cite{DAmico:2019fhj}, we impose the following priors:
For the non-marginalized EFT parameters, we choose non-informative flat priors: $[0, 4]$ on $b_1$ and $[-4, 4]$ on $c_2 = (b_2+b_4)/\sqrt{2}$, and set $b_2-b_4 = 0$ as we find that $b_2$ and $b_4$ are more than 99\% correlated. 
All EFT parameters are normalized such that they are all of the same order $\sim \mathcal{O}(2)$ (same order as $b_1$). 
We thus allow the marginalized EFT parameters to vary only within their physical range: we choose Gaussian priors centered on $0$ of width $2$ on $b_3$, $c_{\rm ct}$, $c_{\epsilon,0}/\bar n_g$, $c_{\epsilon, {\rm quad}}/\bar n_g \equiv \tfrac{2}{3} c_{\epsilon, 2} f /\bar n_g$, and we choose $k_{\rm M}=0.7 \hinvMpc$. 
However, we set $c_{\epsilon, {\rm mono}} /\bar n_g \equiv  (c_{\epsilon, 1} + \tfrac{1}{3}c_{\epsilon, 2} f)/\bar n_g =0$ as we find that the signal-to-noise is too weak to measure this combination.  For the redshift space counterterms we impose a Gaussian prior centered on $0$ of width $8$ on $c_{r,1}$, as we set $c_{r,2}=0$ since it is degenerate with $c_{r,1}$ when only the monopole and the quadrupole are fit. 

When fitting the BOSS data, we use one set of EFT parameters $\lbrace b_1, c_2, b_3, c_{\rm ct}, c_{r,1}, c_{\epsilon, 0}, c_{\epsilon,2} \rbrace$ per skycut.  
{For the fit on $\nu \Lambda$CDM,} in addition to the non-marginalized EFT parameters $b_1$ and $c_2$, we sample over the cosmological parameters $\omega_b$, $\omega_{cdm}$, $h$, $\ln (10^{10}A_s)$, $n_s$ and~$\sum m_\nu$, imposing a BBN prior on $\omega_b$ as discussed in the main text. 
We take the normal hierarchy for the neutrino masses and use a flat prior $0.06 < \sum m_\nu / e\rm{V} < 1$.
When analyzing $w$CDM, we sample over $\omega_b$, $\omega_{cdm}$, $h$, $\ln (10^{10}A_s)$, $n_s$ and $w$ instead, with one massive neutrino fixed to minimal mass $0.06$, with the BBN prior. 

\begin{figure}[h]
\centering
\includegraphics[width=0.99\textwidth]{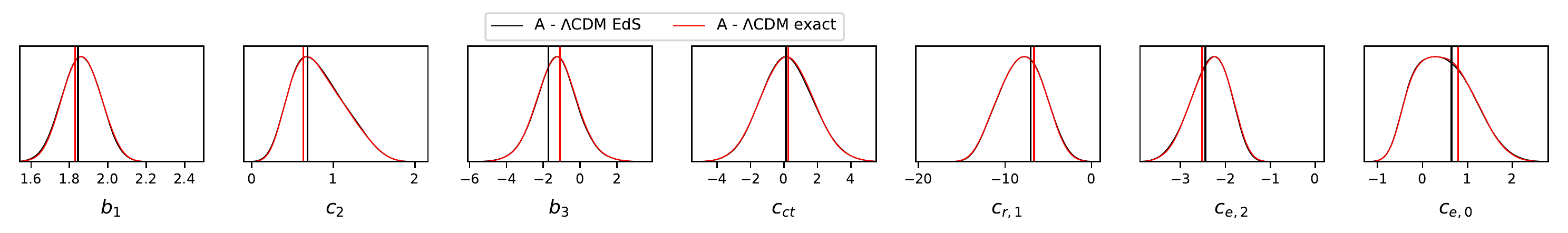}
\includegraphics[width=0.99\textwidth]{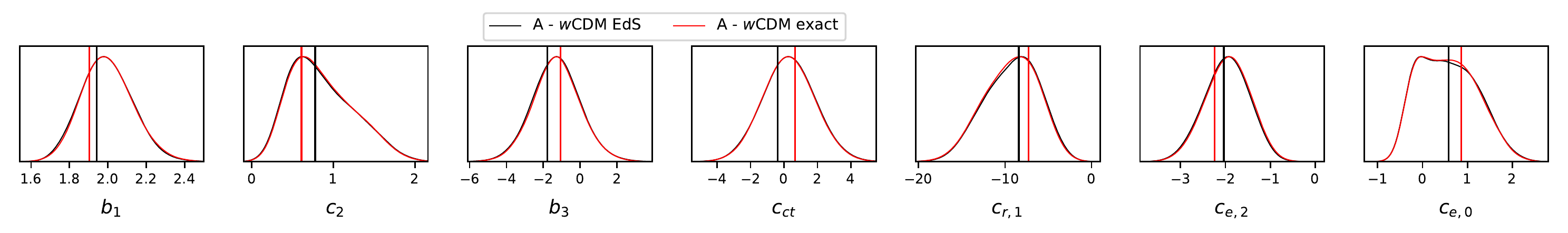}
\includegraphics[width=0.99\textwidth]{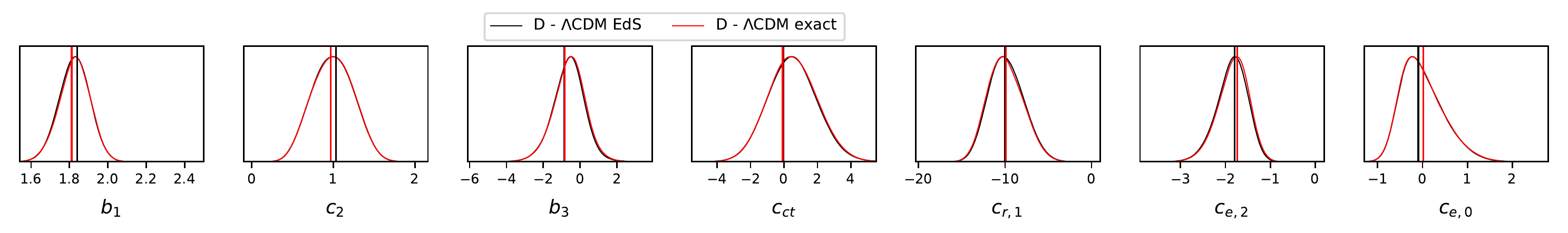}
\includegraphics[width=0.99\textwidth]{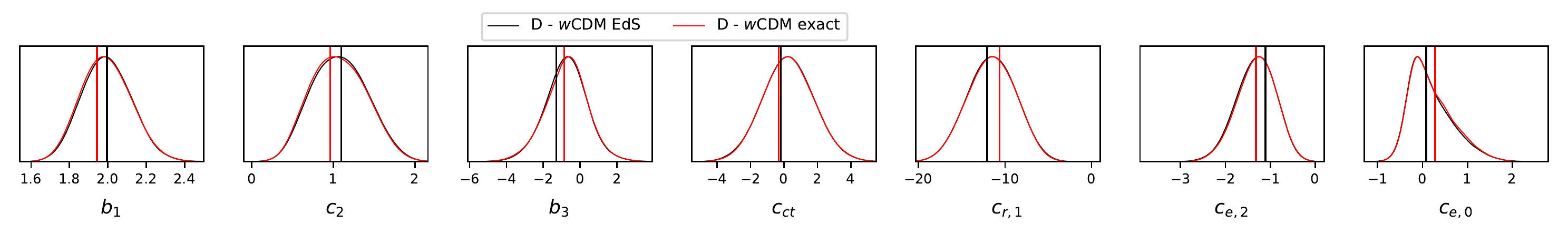}
\caption{EFT parameters obtained by fitting the power spectrum multipoles of the lettered challenge simulations A and D on $\Lambda$CDM and $w$CDM with a BBN prior, with and without the EdS approximation.
The best fit values are depicted as black and red vertical lines, respectively. 
Here we show the posteriors for $b_1$ and $c_2$ obtained using the partially-marginalized likelihood, Eq.~\eqref{eq:margP}, while for the other EFT parameters that are marginalized over at the level of the likelihood, are shown the {\it profiled posteriors} obtained from maximizing the non-marginalized likelihood as given by Eq.~\eqref{eq:bg}, at each point in the MCMC chain sampled using the partially-marginalized likelihood.
} 
\label{fig:challenge_eftparam}
\end{figure}

\paragraph{EFT parameters} In Fig.~\ref{fig:challenge_eftparam}, we show the EFT parameters obtained fitting the power spectrum multipoles of the lettered challenge simulations, with and without the EdS approximation. 
We see that the best fit values are shifted when exact time dependence is accounted for, but by a small amount compared to the error bars.  
However, the 68\% and 95\% confidence levels barely change, as found for the cosmological parameters presented in the main text.
This shows that the EdS time approximation can be used without biasing contours on data volume up to $50$Gpc${}^3$.

\bibliographystyle{JHEP}
\bibliography{references}

\end{document}